\documentclass[12pt,preprint]{aastex}

\def\rmd{{\rm d}}
\def\cqg{Class. Quantum Grav.~}
\newcommand{\erf}[1]{(\ref{#1})}
\newcommand{\Msun}{\ensuremath{M_{\odot}}}
\newcommand{\bel}[1]{\begin{equation}\label{#1}}
\newcommand{\ee}{\end{equation}}

\shorttitle{Intermediate-Mass-Ratio Inspirals for AdvLIGO}
\shortauthors{Mandel, Brown, Gair, Miller}

\begin{document}

\title{Rates and Characteristics of Intermediate Mass Ratio Inspirals
Detectable by Advanced LIGO}

\author{Ilya Mandel}
\affil{Theoretical Astrophysics, California
Institute of Technology, Pasadena, CA 91125;\\
Department of Physics and Astronomy, Northwestern University, 
Evanston, IL 60208} 
\email{ilyamandel@chgk.info}

\author{Duncan A. Brown} \affil{LIGO Laboratory, California Institute of
Technology, Pasadena, CA 91125;\\
Theoretical Astrophysics, California Institute of Technology,
Pasadena, CA 91125;\\
Department of Physics, Syracuse University, Syracuse, NY 13244}

\author{Jonathan R. Gair} \affil{Institute of Astronomy, Madingley Road,
Cambridge, CB3 0HA, UK}

\and

\author{M. Coleman Miller} \affil{University of Maryland, Department of
Astronomy, College Park, MD 20742}

\begin{abstract}

Gravitational waves (GWs) from the inspiral of a neutron star (NS) 
or stellar-mass black hole (BH) into an intermediate-mass black hole 
(IMBH) with mass $M\sim 50 M_\odot$ to $350 M_\odot$ may be 
detectable by the planned advanced generation of ground-based GW 
interferometers. Such intermediate mass ratio inspirals (IMRIs) are 
most likely to be found in globular clusters. We analyze four 
possible IMRI formation mechanisms: (1) hardening of an NS--IMBH or 
BH--IMBH binary via three-body interactions, (2) hardening via Kozai 
resonance in a hierarchical triple system, (3) direct capture, and 
(4) inspiral of a CO from a tidally captured main-sequence star; we 
also discuss tidal effects when the inspiraling object is an NS. For 
each mechanism we predict the typical eccentricities of the 
resulting IMRIs. We find that IMRIs will have largely circularized 
by the time they enter the sensitivity band of ground-based 
detectors. Hardening of a binary via three-body interactions, which 
is likely to be the dominant mechanism for IMRI formation, yields 
eccentricities under $10^{-4}$ when the GW frequency reaches 
$10$~Hz.  Even among IMRIs formed via direct captures, which can 
have the highest eccentricities, around $90\%$ will circularize to 
eccentricities under $0.1$ before the GW frequency reaches $10$~Hz. 
We estimate the rate of IMRI coalescences in globular clusters and 
the sensitivity of a network of three Advanced LIGO detectors to the 
resulting GWs. We show that this detector network may see up to tens 
of IMRIs per year, although rates of one to a few per year may be 
more plausible. We also estimate the loss in signal-to-noise ratio 
that will result from using circular IMRI templates for data 
analysis and find that, for the eccentricities we expect, this loss 
is negligible.

\end{abstract}


\maketitle

\section{Introduction}

Observational evidence from cluster dynamics and from ultra-luminous 
X-ray sources suggests that there may exist a population of 
intermediate-mass black holes (IMBHs) with masses in the $M\sim 
10^{2} - 10^{4}\,M_\odot$ range~\citep{MC04, Trenti06b}.  Numerical 
simulations of globular clusters suggest that IMBHs could merge with 
numerous lower-mass compact objects (COs) during the lifetime of the 
cluster \citep{Tan00,MH02a,MH02b,MT02a,MT02b,GMH04,GMH06,OL06,OL07}, 
through a combination of emission of gravitational radiation, binary 
exchange processes, and secular evolution of hierarchical triple 
systems. Gravitational waves (GWs) will be generated during the 
intermediate-mass-ratio inspiral (IMRI) of a stellar-mass object 
(black hole [BH] or neutron star [NS], since a white dwarf or a 
main-sequence star would be tidally disrupted) into an IMBH.  For 
IMBH mass $\lesssim 350\,M_\odot$, these waves are potentially 
detectable with the planned advanced generation of ground-based GW 
interferometers: Advanced LIGO and its international 
partners~\citep{Barish:1999,Fritschel}.

IMRIs will be important as probes of strong gravity and cluster 
dynamics due to their mass range and dynamical histories. For 
example, from Advanced LIGO IMRI data it may be possible to measure 
the quadrupole moment, $Q$, of an IMBH to an accuracy of $\Delta Q 
\sim Q_{\rm Kerr}$, where $Q_{\rm Kerr}$ is the quadrupole moment of 
a Kerr BH~\citep{Brown}. This is sufficient to distinguish a BH from 
a boson star, for which the quadrupole moment can be many times the 
Kerr value. In addition, since the formation of IMBHs in clusters 
seems to require short mass segregation timescales (see 
\S~\ref{sec:astro}), detection of IMBH mergers and their associated 
masses will yield information about young dense clusters and their 
evolution.

In this paper we discuss the astrophysical and data analysis aspects 
of IMRIs. In \S~\ref{sec:astro} we provide the astrophysical 
setting for IMRIs and describe the formation mechanisms.  We 
estimate the typical eccentricities resulting from different capture 
mechanisms and find that inspirals will largely circularize by the 
time the GW frequency reaches the Advanced LIGO band ($f_{\rm GW} 
\gtrsim 10$ Hz). We show, in particular, that three-body hardening, 
which is likely to be the dominant IMRI formation mechanism, will 
result in binary eccentricities $e<10^{-4}$ in the Advanced LIGO 
band.  Even direct capture, which is the most likely mechanism to 
yield high eccentricities, leads to $\sim 90\%$ of IMRIs with 
$e<0.1$ in the Advanced LIGO band. In \S~\ref{sec:rates} we 
estimate an upper limit on the rate of IMRIs detectable by Advanced 
LIGO of up to $10$ events per year.  A more sophisticated, but 
model-dependent, rate estimate ranges from one event per $3$ years 
for NS IMRIs to $10$ events per year for $10\ M_\odot$ BH IMRIs.  
The event rate can be enhanced by a factor of $\sim 3.5$ by 
optimizing Advanced LIGO for detections at low 
frequencies.\footnote{We used the Advanced LIGO Bench code to 
perform this optimization: http://www.ligo.mit.edu/bench/bench.html} 
Searches for IMRIs in Advanced LIGO data will likely use matched 
filtering techniques, for which accurate waveform templates are 
required. In \S~\ref{sec:ecc}, we estimate that there will be a 
negligible loss in signal-to-noise ratio (S/N) if circular templates are 
used to search for IMRIs with the expected eccentricities in 
Advanced LIGO data.

\section{Astrophysical Setting, Capture Mechanisms,
and Typical Eccentricities}\label{sec:astro}

An IMBH cannot result from the evolution of a solitary star in the
current universe because even a star of initial mass $\sim
10^2\,M_\odot$ will be reduced well below this mass by winds and
pulsational instabilities driven by metal-line opacities
(cf.~\cite{FK2001},
Fig.~7 and associated discussion). Some IMBHs might be formed from
the first, metal-free, stars \citep{MR01}, but these IMBHs are unlikely
to participate in multiple mergers with COs. Instead, we
focus on the proposal that IMBHs can be produced in the current universe
via runaway stellar collisions in dense young stellar clusters.  If the
most massive stars segregate to the center in less than their $\sim
2\times 10^6$~yr lifetimes
\citep{Ebisuzaki01,PM02,PZ04,GFR04,GFR06,FRB06,FGR06,Fregeau06}, then
stellar mergers can overcome mass losses and the collision product can
reach hundreds to thousands of solar masses, presumably evolving into an
IMBH.

When supernovae start to occur, the resulting mass loss leads to an
expansion
of the cluster, which thus transitions into a more collisionless stage of
existence.  From this point on, COs can be captured by the IMBH and
generate
observable GWs as they inspiral under radiation reaction
and
eventually merge with the IMBH.

Early in the history of the globular cluster the inspiraling objects in
IMRIs are likely to be $m\sim 10\ \Msun$ BHs, which may form a
dense subcluster composed purely of BHs around the
IMBH~\citep{OL06, OL07}.  Late in the cluster's history, once the BH
central subcluster has largely evaporated, NSs will
likely replace the larger BHs as the inspiraling objects.

There are several ways in which stellar-mass COs can be captured by an
IMBH. Most mechanisms of capture involve binaries, because the cross
section of a binary is orders of magnitude larger than that of a single
CO.

Extensive numerical studies of binary-single interactions demonstrate
that hard binaries (defined, e.g., so that the total energy of the
binary-single system is negative) tend to be tightened by three-body
interactions \citep{Heggie75}.  These studies also show that massive
objects such as stellar-mass BHs and IMBHs tend to swap into
binaries.  The most likely capture mechanism involves the formation of
a CO--IMBH binary, which is subsequently hardened by repeated three-body
interactions until radiation reaction becomes significant and the binary
coalesces.

Hardening can also occur via binary-binary interactions; unlike
binary-single interactions, these can result in a stable hierarchical
triple. Some fraction of these end up in orientations favorable for the
secular Kozai resonance \citep{K62}, in which the inner binary (which
contains the IMBH) periodically increases and decreases its eccentricity
while keeping its semimajor axis constant.  The periapsis distance can
therefore be quite low in parts of the cycle and can lead to
coalescence without Newtonian recoil \citep{MH02b,Wen02}, although
recoil from gravitational radiation will still occur (see
\S~\ref{IMBHdensity}).  The importance of the Kozai resonance depends
on the frequency of binary-binary interactions, which is unknown at
present.

In addition to these mechanisms, which usually require multiple
interactions to lead to merger, a hyperbolic encounter at a close
enough periapsis can produce direct capture via emission of
gravitational radiation.  Assuming that the IMBH does not have a
significant radius of influence, the effective cross section for
radiative capture of the CO by an IMBH is proportional to
$M^{12/7}$, where $M$ is the mass of the IMBH~\citep{QS87}.   For
two-body encounters this process is likely to be important only for
masses high enough ($\gtrsim 10^3\,M_\odot$) that the frequency of
the GWs throughout the subsequent inspiral will be
below the sensitivity range of ground-based detectors.  However,
direct capture during a three-body interaction could be
significant~\citep{GMH06}.

Finally, an IMBH could tidally capture a main-sequence star. If the
captured star evolves to a CO while in orbit around the
IMBH, the remnant could remain bound to the IMBH and ultimately spiral
in via GW emission. This scenario has been suggested as
a possible explanation for the observed population of ultraluminous
X-ray sources~\citep{HPZA03,HPZ05}.

Additionally, orbital energy may couple to vibrational normal modes of
the inspiraling object in the case when the inspiraling object is an
NS.  In principle, the energy loss to tidal heating of an
NS could change the inspiral trajectory, or even disrupt the
NS.

The IMRI enters the Advanced LIGO band when
\bel{fGW}
f_{\rm GW}=\frac{\omega_{\rm orb}(r_p)}{\pi}
        = \frac{1}{\pi} \sqrt{\frac{M}{r_p^3}} \gtrsim 10~{\rm Hz},
\ee
i.e., when the periapsis is $r_p \approx 16 GM/c^2 = 1600 G M_\odot/c^2$
for
$M=100  M_\odot$.  The frequency of the dominant quadrupolar ($k=2$)
harmonic in the GWs emitted at the innermost stable
circular orbit is
\bel{fISCO}
f_{\rm GW,\ ISCO} \approx 44.0 \,\frac{M}{100\ M_\odot}~{\rm Hz}
\ee
for inspirals into non-spinning BHs.

Below, we discuss the eccentricity of an IMRI at the time its
GW frequency enters the Advanced LIGO band for each of
the mechanisms mentioned above:  (1) formation of a CO--IMBH binary and
subsequent hardening via three-body interactions, (2) Kozai resonance
of a hierarchical triple system, (3) direct capture when a solitary CO
passes close to the IMBH, and (4) tidal capture of a main-sequence star
that subsequently evolves to leave a CO.  We also consider
the impact of (5) tidal interactions with an inspiraling NS.

\subsection{Hardening of a CO--IMBH Binary via Three-Body Interactions}
\label{harden}

This mechanism proceeds as follows. The IMBH rapidly swaps into a binary
because it is far more massive than any other object in the globular
cluster.
Advanced LIGO can detect IMRIs at redshifts $z \lesssim 0.2$
(\S~\ref{LIGOrange}), so it will predominantly observe globular clusters
late
in their history.  On a timescale that is short relative to the merger
time,
an NS or a BH will encounter the binary containing the
IMBH
and will exchange for the companion in this binary, since stellar
remnants are
the most massive objects in the late cluster other than the IMBH itself.
From
that point on, stars of all types (although biased towards the heavy ones
that
segregate towards the center) engage in three-body interactions.
Numerical
simulations show that interactions tend to tighten a binary if it is hard.
This can be understood heuristically for three equal-mass objects by
noting
that ejection will tend to occur at roughly the binary orbital speed;
hence, if
this is greater than the initial encounter speed at infinity the binary
loses
energy.  Binary tightening proceeds until the binary can merge through
radiation reaction from the emission of GWs.

For the unequal mass binaries we consider here, simulations by
\citet{Q96} show that a single interaction of a field star of
mass $m_*$ with a binary of total mass $M$ will on average change the
binary energy by a fractional amount $\Delta E/E={\cal O}(m_*/M)$,
roughly independent of the component masses of the binary.  More
precisely,
approximately $(2 \pi /22)  M / m_*$ interactions are required to reduce
the
semimajor axis of a hard binary by one $e$-folding \citep{Q96}.

The rate at which objects interact with the IMBH binary is
\bel{intrate}
{\dot N}=n\varsigma v,
\ee
where $n$ is the number density, $v$ is the relative speed, and
$\varsigma$ is the gravitationally focused cross section $\varsigma=\pi
a(2GM/v^2)$ for an interloper to approach within the binary's semimajor
axis $a$ of the binary. Since this rate is proportional to $a$, the
total time for the binary to harden until the semimajor axis equals $a$
is dominated by the last $e$-folding time:
\bel{Tharden}
T_{\rm harden} \approx \frac{2 \pi}{22} \frac{M}{m_*} \frac{1}{\dot N}
        \approx 2 \times 10^8
         \left(\frac{10^{13}\ {\rm cm}}{a}\right)\ {\rm yr},
\ee
where we set $m_*=0.5\ M_\odot$, $v=10^6$~cm~s$^{-1}$, and
$n=10^{5.5}$~pc$^{-3}$ (the number density of all stars in a
core-collapsed globular cluster; \citealt{pryor93}).

For a sufficiently massive BH, a cusp is formed and the
   interactions are no longer described by individual binary-single
   encounters.  We can estimate roughly the mass above which this
   occurs.  Consider a core of number density $n_{\rm core}$ and
   velocity dispersion $\sigma$.  For an IMBH of mass $M$, the
   radius of influence (inside of which the IMBH dominates the
   potential) is $r_{\rm infl}=GM/\sigma^2$.  For a true cusp,
   \cite{BahcallWolf} showed that the number density
   would go as $n(r)=n_{\rm core}(r/r_{\rm infl})^{-7/4}$.  The
   total number of objects inside $r_{\rm infl}$ is then
   \begin{equation}
   N(r<r_{\rm infl})=\int_0^{r_{\rm infl}}4\pi r^2n_{\rm core}
   (r/r_{\rm infl})^{-7/4}dr=(16\pi/5)r_{\rm infl}^3n_{\rm core}\; .
   \end{equation}
   Scaling to canonical values, this gives
   \begin{equation}
   N(r<r_{\rm infl})\approx 0.3(M/100\,M_\odot)^3
   (\sigma/10~{\rm km~s}^{-1})^{-6}(n_{\rm core}/10^{5.5}~{\rm pc}^{-3})\;
\end{equation}
   Therefore, in the mass range most relevant to Advanced LIGO,
   it is unlikely that there will be a significant cusp, hence our
   treatment of isolated binary-single interactions is reasonable.
   For more massive BHs a cusp might form, although we note
   that for $M<1000\,M_\odot$ the typical distance wandered by the
   IMBH in the core is larger than the radius of influence; hence,
   cusp formation could be made more difficult.  This is, however,
   worth further study.

The gravitational radiation merger timescale for an
intermediate-mass-ratio binary of semimajor axis $a$,
eccentricity $e$, reduced mass $\mu \approx m$,
and total mass approximately equal to the IMBH mass $M$, is \citep{P64}
\bel{Tmerge}
T_{\rm merge} \approx 10^{17} \frac{\Msun^3}{M^2 m}
        \left(\frac{a}{10^{13}\ {\rm cm}}\right)^4 (1-e^2)^{7/2}\ {\rm
yr}.
\ee

Simulations and phase-space arguments show that three-body interactions
cause the eccentricity of the binary to sample a thermal distribution
$P(e)de=2 e de$~\citep{Heggie75} in the Newtonian realm where
gravitational radiation is not significant. If an interaction leaves the
binary with a high eccentricity, however, it is more likely to merge.
\citet{GMH06} examined the eccentricity of the binary
after its final three-body encounter and found a typical value of $e
\approx 0.98$ due to this bias.  Taking this as the typical value for
eccentricity, we find
\bel{Tmerge2}
T_{\rm merge} \approx 10^{8}\left(\frac{M_\odot}{m}\right)
         \left(\frac{100\ M_\odot}{M}\right)^2
        \left(\frac{a}{10^{13}\ {\rm cm}}\right)^4\ {\rm yr}.
\ee
In fact, there is a distribution of eccentricities after the last encounter, rather than a single eccentricity value of $0.98$.  However, Monte Carlo simulations, which are described below, confirm the typical merger times and final eccentricities computed here analytically by assuming the final-encounter eccentricity of $0.98$.  Moreover, simulations indicate that the fraction of direct plunges from highly radial orbits must be extremely small, because they would require improbably small periapsis separations.

The IMRI rate will be maximized when the total merger time, $T = 
T_{\rm harden}+T_{\rm merge}$, is minimized.  Minimizing $T$ with 
respect to $a$, we find that the total merger time is $T \approx 3 
\times 10^8\ {\rm yr}$ for the inspiral of an $m=1.4\ M_\odot$ NS 
into an $M=100\ M_\odot$ IMBH, yielding an IMRI rate per globular 
cluster of $\sim 3 \times 10^{-9}$~yr$^{-1}$.  If the inspiraling 
object is an $m=10\ M_\odot$ BH, and the IMBH mass is again $M=100\ 
M_\odot$, then the total merger time is $T \approx 2 \times 10^8\ 
{\rm yr}$, and the IMRI rate per globular cluster is $\sim 5 \times 
10^{-9}$~yr$^{-1}$.

These numbers are close to the answers obtained with Monte Carlo 
simulations using the same procedure as in G\"ultekin, Miller, \& 
Hamilton (2006).  We find from these simulations that the total time 
to merger averages $5\times 10^8$~yr for $1.4~M_\odot$ NSs and 
$3\times 10^8$~yr for $10\,M_\odot$ BHs, interacting with field 
stars of mass $0.5~M_\odot$ and an IMBH of mass $100~M_\odot$.  We 
also find that, as we assumed, once a CO is in a binary with an IMBH 
it stays there; only a fraction $\approx 6\times 10^{-4}$ of 
encounters swapped out an NS, and only 1 of the $5\times 10^4$ 
encounters swapped out a BH.  Therefore, as we indicated, the object 
that eventually merges with the IMBH is highly likely to be a CO.

This mechanism requires the cluster to be in a core-collapsed state 
and for this state to persist for a time $\gg 2 \times 10^8$yr. Core 
collapse is expected to persist in the absence of significant 
heating, as will be the case for clusters with IMBHs in the mass 
range of interest, so $2 \times 10^8$yr should be easily achievable. 
About $20\%$ of clusters currently are in a state of core collapse, 
so this state can indeed be sustained for times of order a Hubble 
time, or much longer than $2 \times 10^8$yr.  We consider only these 
core-collapsed systems as likely hosts of IMRIs when computing event 
rates below.

Radiation reaction from GW emission  dominates the
evolution once the GW merger timescale $T_{\rm merge}$
[Eq.~(\ref{Tmerge2})] is
shorter than the average time between three-body encounters, $1/\dot{N}$,
defined by Eq.~\ref{intrate}.  For the NS--IMBH system ($m=1.4\ \Msun$,
$M=100\
\Msun$), this occurs when the semimajor axis takes the value $a \approx
5\times10^{12}$~cm.  As discussed earlier, the eccentricity at this time
is
$e\approx 0.98$, and hence the periapsis is $r_p \approx
10^{11}$~cm~$\approx
7000\ GM/c^2$.  For the BH--IMBH system ($m=10\ \Msun$, $M=100\ \Msun$),
radiation reaction dominates for $a \lesssim 8\times10^{12}$~cm,
corresponding
to a periapsis of $r_p \approx 1.6 \times 10^{11}$~cm~$\approx 10000\
GM/c^2$.

Keplerian orbits evolving under radiation reaction satisfy [see
Eq.~(5.11)
of~\cite{P64}]
\bel{erp}
         r_p e^{-12/19} (1+e) \left[1+(121/304) e^2\right]^{-870/2299}
         = {\rm  constant},
\ee
from which we can obtain the eccentricity at a particular frequency, given
the
initial values of periapsis and eccentricity.  We find that for this
capture
mechanism, the eccentricity when the source enters the Advanced LIGO band
($f_{\rm GW}=10$ Hz) is very small: $e\lesssim 3\times 10^{-5}$ for the
NS--IMBH system and $e\lesssim 2\times 10^{-5}$ for the BH--IMBH system.
The
orbit will thus have circularized by the time the IMRI is in the Advanced
LIGO band. This is consistent with the results of~\cite{GMH06}, who also found that IMRI binaries formed through this channel would circularize before they entered the Advanced LIGO band.

\subsection{Kozai Resonance}

A stable hierarchical triple system could experience Kozai resonance 
that would drive the eccentricity of the inner binary to a value 
close to unity \citep{K62}, leading to a small periapsis separation 
and binary tightening and eventual merger through gravitational 
radiation reaction \citep{MH02b, Wen02}.  Some simulations (e.g., 
those of \citet{OL06}) suggest that the four-body (binary-binary) 
interactions that are required to place the binary on the Kozai 
merger track constitute only a small fraction of the total number of 
merger events in the cluster.  If so, four-body interactions play a 
minor role in IMRI formation.  These simulations may not consider 
all possibilities, however.  In particular, in the \cite{OL06} 
model, binaries are only destroyed (through mergers, or by being 
kicked out of the subcluster).  Therefore, the binary fraction 
decreases with time, meaning that binary-binary interactions are 
uncommon late in the cluster's history.  There may be a way to 
replenish the number of BHs in binaries, however.  Approximately 
$5\%$--$20\%$ of normal stars in globulars are in binaries 
\citep{rubenstein97, bellazzini02} (this fraction is closer to 
$50\%$--$70\%$ in the field, but in globulars the wide binaries are 
disrupted).  If such a binary goes through the BH subcluster, a BH 
could swap in, so that even if no BHs were originally in binaries, 
the binary fraction could increase.

Although computing the relative contribution of Kozai resonance 
mergers to the total number of IMRIs requires more detailed modeling 
of the cluster dynamics, it is possible to estimate the largest 
eccentricity that could result from this mechanism (see \cite{Wen02} 
for a more detailed discussion in the context of stellar-mass BHs).  
For this calculation, we assume that the Kozai resonance drives the 
binary to a sufficiently high eccentricity to allow merger via 
radiation reaction within one Kozai cycle. In reality, the semimajor 
axis and eccentricity would evolve gradually over multiple Kozai 
cycles, leading to larger typical periapses and smaller 
eccentricities, so our assumption will overestimate the typical 
eccentricities of IMRIs in the Advanced LIGO band.

We assume that the eccentricity is near its maximum for a fraction 
$0.01$ of the total Kozai cycle (based on Fig.~1 of \cite{I97}) and 
compare this time with the radiation reaction timescale.  If the 
radiation reaction merger time is much longer than the time near 
maximum eccentricity, we assume that gravitational radiation is 
insignificant.  If instead the timescale for Kozai resonance to 
drive the eccentricity to some value $e\approx 1$ is much larger 
than the timescale for radiation reaction to circularize the orbit 
down from $e$, then the eccentricity will never reach $e$ in 
practice, even though $e$ may be less than the maximum possible 
eccentricity for the given configuration (see below). Therefore, the 
maximum eccentricity reachable when including gravitational 
radiation is given approximately by the condition that the radiation 
reaction timescale is equal to the time near that high eccentricity.

The timescale for the Kozai cycle is given by, e.g.,  Eq.~(4) of
\cite{MH02b}:
\bel{Kozai}
\tau_{\rm Kozai}\approx {\rm few} \times
         \left( \frac{M_1 b_2^3}{m_2 a_1^3}\right)^{1/2}
         \left(\frac{b_2^3}{G m_2}\right)^{1/2} \approx
         3 \times \left(\frac{M_1}{100\ M_\odot}\right)^{1/2}
         \left(\frac{M_\odot}{m_2}\right) \left(\frac{b_2}{a_1}\right)^3
         \left(\frac{a_1}{10^{13}\ {\rm cm}}\right)^{3/2}\ {\rm yr},
\ee
where, in the notation of \cite{MH02b}, $M_1$ is the total mass
(approximately
equal to the mass of the IMBH), $m_1$ is the mass of the inner 
companion, $m_2$ is the mass of the outer companion,
$a_1$
is the semimajor axis of the inner binary, and $b_2$ is the semiminor axis
of
the outer binary.   Setting the timescale for merger by gravitational
radiation, given in  Eq.~(\ref{Tmerge}), equal to $\tau_{\rm GR}=0.01\
\tau_{\rm Kozai}$ yields
\bel{epsilon}
         \left(\frac{a_1}{10^{13}\ {\rm cm}}\right)^{5/2} \epsilon^{7/2}
\approx
         3 \times 10^{-15} \left(\frac{M_1}{100\ M_\odot}\right)^{5/2} \, 
	\left(\frac{m_1}{m_2}\right) \, \left(\frac{b_2}{a_1}\right)^3
\ee
where $\epsilon \equiv 1-e^2$.

Relativistic precession constrains the maximal eccentricity, or minimal
$\epsilon$, that can be achieved in a Kozai cycle.  That minimal
$\epsilon$ is
given by Eqs.~(6) and (8) of \cite{MH02b} as:
\bel{minepsilon}
         \epsilon \approx \frac{1}{9}
         \left(8 \frac{b_2^3 G M_1^2}{m_2 a_1^4 c^2}\right)^2
         \approx 1.6 \times 10^{-7}
         \left(\frac{m_2}{M_\odot}\right)^{-2}
         \left(\frac{M_1}{100\ M_\odot}\right)^4
         \left(\frac{a_1}{10^{13}\ {\rm cm}}\right)^{-2}
\left(\frac{b_2}{a_1}\right)^6.
\ee

In order to compute the maximal plausible eccentricity at $f_{\rm GW}=10$
Hz,
we need to estimate the minimal periapsis radius at the peak of the Kozai
cycle,
when radiation reaction becomes dominant, since eccentricity will be close
to
unity there [cf.~Eq.~(\ref{erp})].  That is, we must minimize
$r_p = a_1 (1-e) \approx a_1 \epsilon /2$.  This minimum value is found by 
solving Eqs.~(\ref{epsilon}) and (\ref{minepsilon}). We find
\bel{minaeps}
         \frac{a_1 \epsilon}{10^{13}\ {\rm cm}} \gtrsim 1.8 \times 10^{-5}
         \left(\frac{m_2}{M_\odot}\right)^{-4/9}
         \left(\frac{M_1}{100\ M_\odot}\right)^{13/9}
         \left(\frac{b_2}{a_1}\right)^2
	 \left(\frac{m_1}{m_2}\right)^{2/9}.
\ee

Stability requires that the semiminor axis of the outer binary is at least
a
few times greater than the semimajor axis of the inner binary, so we set
$b_2/a_1=5$.  We again assume $M_1 = 100\ M_\odot$, $m_1=1.4\ M_\odot$, 
and $m_2=M_\odot$
(although
this choice violates the restricted three-body assumption under which
Eq.~(8)
of \citep{MH02b} was derived).  These parameter values predict a minimal
$r_p
\gtrsim 170 G M/c^2$ at the time when radiation reaction takes over;
hence,
according to Eq.~(\ref{erp}), the maximal eccentricity of IMRIs formed via
the
Kozai resonance mechanism in the Advanced LIGO band is $e \approx 0.01$.

\subsection{Direct Captures}
\label{dircapt}

If we assume that the IMBH is wandering in the stellar cluster, the
effective
cross section for direct captures via two-body relaxation
(GW emission) is proportional to the $(12/7)$ power of the total mass
\citep{QS87},
so an IMBH has
a relatively small capture cross section, making this capture mechanism
relatively unlikely. If we instead assume that the $M-\sigma$ relation
holds
for globular clusters, which is equivalent to saying that the IMBH
dominates
the dynamics in the center of the cluster, the capture rate would increase
towards smaller IMBH masses, like $M^{-1/4}$~\citep{HA05}, and this
channel
would contribute significantly to the total rate. However, as discussed in 
\S~\ref{harden}, the IMBHs of
interest for Advanced LIGO, with $M\sim 100\Msun$, have a very small
radius of
influence and so they will not have a significant influence on the
dynamics in
the cluster center.  The direct capture mechanism, in any case,
can yield higher eccentricities than scenarios involving binaries.

The critical periapsis separation $r_p$ for the direct capture of a
CO of mass $m$, moving at infinity with velocity $v$, by an IMBH of
mass $M
\gg m$ is [e.g.~Eq.~(11) of~\cite{QS89}]:
\bel{rpcapture}
\frac{r_p^{\rm max} c^2}{GM} \approx 950
         \left(\frac{m}{M}\right)^{2/7}
         \left(\frac{v}{10^6\ {\rm cm\ s^{-1}}}\right)^{-4/7}.
\ee
If $M=100\ M_\odot$, $m=1.4\ M_\odot$, and $v=10^6$ cm s$^{-1}$, direct
capture
is possible at a maximum periapsis of $r_p^{\rm max} c^2/(GM) \approx
280$; if
$m=10\ M_\odot$ and $M$ and $v$ are the same as above, the maximum
periapsis is
$r_p^{\rm max} c^2/(GM) \approx 500$.  For such small periapses,
gravitational
focusing implies $r_p \propto b^2$, where $b$ is the impact parameter.
Hence,
the probability distribution $P(b) \propto b$ in impact parameter
corresponds
to a uniform distribution in periapsis at capture, $P(r_p) = {\rm
constant}$.

\begin{figure}
\plotone{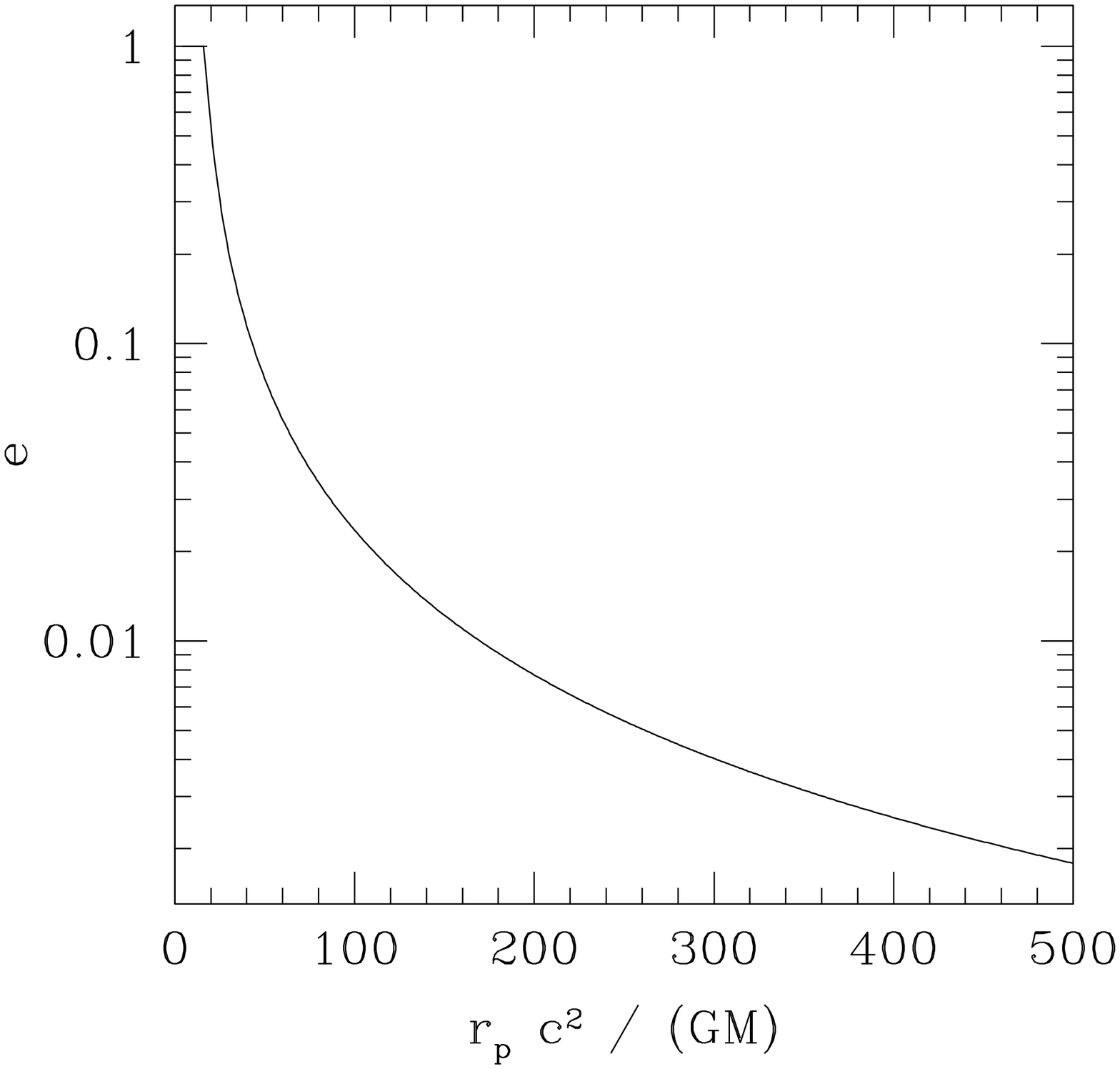}
\figcaption{Eccentricity at $f_{\rm GW}=10$ Hz as a
function
of the periapsis at capture, for a CO inspiraling into an
IMBH of mass $M=100\ M_\odot$.  The eccentricity at capture is set to $1$,
and the eccentricity at $r_p \approx 16\ GM/c^2$, where $f_{\rm GW}=10$
Hz,
follows from Eq.~(\ref{erp}).
\label{eccdist}}
\end{figure}

In Figure~\ref{eccdist}, we plot the eccentricity of an IMRI at the
frequency
at which it enters the Advanced LIGO band as a function of the initial
periapsis at capture, following Eq.~(\ref{erp}).  The initial eccentricity
at
capture can be computed from the energy lost during the first pass;
however,
the exact value does not significantly affect the eccentricity at $f_{\rm
GW}=10$ Hz, so we
set the eccentricity at capture to be $e=1$.  The initial periapsis is
uniformly distributed between $r_p^{\rm min}=4 GM/c^2$ (orbits with
periapsis
under $4GM/c^2$ will plunge rather than inspiral) and $r_p^{\rm max}$.
Therefore, to determine the total fraction of directly captured IMRIs that
circularize to a given level $e \le e_{\rm cutoff}$ by the time they are
in the
detector band, it is sufficient to find the fraction of the interval
$[r_p^{\rm min}, r_p^{\rm max}]$ for which the line in Fig.~\ref{eccdist}
stays below $e_{\rm cutoff}$.

Thus, for the chosen IMBH mass of $M=100\ M_\odot$, if the CO is an 
$m=1.4\ M_\odot$ NS, $86 \%$ of all directly captured IMRIs will be 
circularized to $e\le 0.1$ by the time they are in the Advanced LIGO 
band.  If the CO is an $m=10\ M_\odot$ BH, $92 \%$ of all directly 
captured IMRIs will be circularized to $e\le 0.1$ and $67 \%$ will 
be circularized to $e \le 0.01$ by the time they are in the detector 
band.

\subsection{Tidal Capture of a Main-Sequence Star}

It has been suggested that ultraluminous X-ray (ULX) sources are 
systems in which a main-sequence star that has been tidally captured 
is transferring mass to an IMBH via Roche lobe 
overflow~\citep{HPZA03}. In such a system, after the star reaches 
the end of its main-sequence lifetime and undergoes a supernova, it 
may leave a CO on an orbit about the IMBH~\citep{HPZ05} and this 
object may then spiral into the IMBH via GW emission. Although work 
on this problem has focused on sources that might be detected by 
LISA, results have also been presented for the $\sim 100\Msun$ IMBHs 
that we consider here. For $M\sim 100\Msun$, only $1\%$--$2\%$ of 
systems leave a CO that inspirals into the IMBH within a Hubble 
time, and these remnants are always NSs~\citep{HPZ05}. Following 
\cite{HPZ05} we can estimate the rate of these events by assuming 
that there is $\sim 1$ ULX source in each galaxy. The ULX phase lasts 
approximately the main-sequence lifetime of the captured star, which 
is $\sim 10^7$ yr, so we estimate that the capture rate is 
$10^{-7}$ yr$^{-1}$. Multiplying by the fraction of events that 
successfully inspiral, we estimate a rate of $1$--$2\times 10^{-9}$ 
IMRIs per galaxy per year. There are typically $\sim 100$ globular 
clusters per galaxy, so the rate per globular cluster is $\sim 
10^{-11}$ yr$^{-1}$, which is considerably smaller than the binary 
hardening rate.  Thus, while this channel could lead to some IMRIs 
detectable by Advanced LIGO, the rate is likely to be significantly 
lower than the binary hardening channel.

An NS captured via this mechanism would begin to inspiral into the 
IMBH with eccentricity $e\lesssim 0.9$~\citep{HPZ05} and periapsis 
approximately equal to the tidal radius, $(M/M_*)^{1/3} R_*$, where 
$M_* \gtrsim 10\Msun$ and $R_*$ are, respectively, the mass and radius 
of the main-sequence star. Assuming, conservatively, $R_* \gtrsim 
10^5$km, this capture periapsis is typically $\gtrsim 500 (GM/c^2)$. 
For an $M=100\Msun$ IMBH, equation~\erf{erp} predicts $e \approx 
0.002$ when the source enters the Advanced LIGO band. In practice, 
the eccentricity is likely to be even smaller. It is thus quite 
clear that this capture mechanism also produces sources that are 
essentially circular when they enter the Advanced LIGO band.

\subsection{Tidal Effects}

If the inspiraling object is an NS, tides may be significantly 
excited as it passes the central IMBH. If sufficient energy goes 
into tidal heating, the NS could be disrupted. Prior to disruption 
the orbital inspiral will be modified as orbital energy and angular 
momentum are lost into tidal heating.  Tidal interactions are not 
important for the IMRI events we are considering, however, as we 
demonstrate below.

\subsubsection{Tidal Disruption}\label{tidaldis}

A star will be tidally disrupted by a BH when the gravitational tidal
force
acting over the star due to the BH exceeds the self-gravity of the star.
Assuming a Newtonian potential, this leads to the usual tidal disruption
radius
\begin{equation}
R_{\rm td} = R_* \,\left(\frac{M}{m}\right)^{\frac{1}{3}}
         = 41.5 {\rm km} \left(\frac{R_*}{10{\rm km}}\right)
         \left( \frac{M/100\Msun}{m/1.4\Msun}\right)^{\frac{1}{3}},
         \label{rtdnewt}
\end{equation}
in which $R_{\rm td}$ is the radius at which tidal disruption occurs, $m$
and
$R_*$ are the mass and radius of the star, respectively, 
and $M$ is the mass of the BH.  The
gravitational field outside a Kerr BH is not Newtonian, but
\erf{rtdnewt} still provides a reasonable estimate of the tidal disruption
radius. Comparing this to the Schwarzschild radius of a $100\Msun$ BH,
$R_{\rm S} = 2GM/c^2 = 300{\rm km}$, suggests that, even when relativistic
effects and BH spin are included, tidal disruption could only
occur very close to the central BH. Earlier in this section we
showed
that the orbits of IMRI objects are effectively circular by the time the
CO
gets close to the IMBH. The tidal effects for stars on circular
orbits are most extreme for prograde equatorial orbits, since these come
closest to the central body. Thus, we use results for prograde,
equatorial circular orbits for a more accurate calculation of tidal
disruption.

\citet{vallis00} analyzed NS disruption using the correct
tidal field for objects in prograde, circular, equatorial orbits around a
Kerr BH and found that the GW frequency at which tidal
disruption occurred, $f_{td}$, satisfied the relationship
\begin{equation}
R_* = \left\{
\begin{array}{lr}3.25{\rm km} \left(m/1.4 M_{\odot}\right)^{\frac{1}{3}}
         \left(M/50 M_{\odot}\right)^{\frac{2}{3}}
         \left(GMf_{\rm td}/c^3\right)^{-0.71}\quad
         &GMf_{\rm td}/c^3\le 0.045\\
         1.55{\rm km} \left(m/1.4 M_{\odot}\right)^{\frac{1}{3}}
         \left(M/50 M_{\odot}\right)^{\frac{2}{3}}
         \left(GMf_{\rm td}/c^3\right)^{-0.95}\quad
         &GMf_{\rm td}/c^3\ge 0.045
         \label{rtdkerr}
\end{array}\right.
\end{equation}
An inspiraling object plunges into the BH when it reaches the innermost
stable
prograde circular orbit (ISCO). This has radius~\citep{BPT72}
\begin{eqnarray}
\frac{c^2 R_{\rm isco}}{GM} &=&
         3+\sqrt{3\chi^2 + Z^2}-\sqrt{(3-Z)(3+Z+2\sqrt{3\chi^2+Z^2})},
         \nonumber \\
\mbox{where } Z &=& 1+\left[(1+\chi)^{\frac{1}{3}} +
         (1-\chi)^{\frac{1}{3}}\right]\left(1-\chi^2\right)^{\frac{1}{3}},
         \label{iscorad}
\end{eqnarray}
where $\chi=S_1/M^2$ is the dimensionless spin parameter of the BH.

The condition that the star is not disrupted before plunge sets a maximum
radius for the NS. If we require the tidal disruption frequency
to be
greater than the frequency of a prograde circular orbit at the ISCO,
$GMf_{\rm
isco}/c^3 = \{\pi [\chi+(c^2 R_{\rm isco}/GM)^{3/2}]\}^{-1}$ ~\citep{BPT72},
then
Eqs.~\erf{rtdkerr}--\erf{iscorad} imply that the NS escapes
disruption provided that
\begin{equation}
R_* < \left\{
\begin{array}{lr}7.33{\rm km} \left(m/1.4 M_{\odot}\right)^{\frac{1}{3}}
         \left(M/50 M_{\odot}\right)^{\frac{2}{3}}
         \left\{\chi+[c^2R_{\rm isco}/(GM)]^{\frac{3}{2}}\right\}^{0.71}\quad
         &\chi\le 0.6894\\
         4.59{\rm km} \left(m/1.4 M_{\odot}\right)^{\frac{1}{3}}
         \left(M/50 M_{\odot}\right)^{\frac{2}{3}}
         \left\{\chi+[c^2R_{\rm isco}/(GM)]^{\frac{3}{2}}\right\}^{0.95}\quad
         &\chi \ge 0.6894
\end{array}\right.
\end{equation}
Reasonable NS models have a maximum radius of $\sim 16$ km or
less, so this criterion will be satisfied for a $50M_{\odot}$ IMBH if the
spin
$\chi<0.95$. For a $100 M_{\odot}$ IMBH, the condition is satisfied for
all
spins up to $0.998$. As discussed later, we expect IMBHs that grow through
minor mergers to have only moderate spin $\chi\lesssim 0.3$, so
tidal disruption should not occur for such IMBHs.

Although the NS cannot be directly tidally disrupted, tidal 
oscillations will be excited every time the star passes through 
periapsis. If sufficient energy is deposited into such tides, the 
star could eventually be disrupted through this tidal 
heating~\citep{freitag02}. To assess whether this effect could be 
important, we consider the orbital energy lost to leave the star on 
an orbit with periapsis $r_p$ and eccentricity $e$ divided by the 
binding energy of the star, $E_{\rm orb}/E_{\rm bind}$. If the 
inspiral was entirely driven by tidal dissipation, and the tidal 
energy was not efficiently radiated, this would be the ratio of the 
energy in tidal oscillations to the stellar binding energy. Under 
these assumptions, if this ratio was of the order of $1$ or more, 
then tidal heating could disrupt the star. In practice, however, 
most of the orbital energy is lost to gravitational radiation, 
since, as we see below, tidal oscillations can only be excited 
during the late stages of inspiral. Thus, most of the energy does not 
go into tidal heating, and therefore this ratio would have to be 
significantly greater than $1$ for tidal disruption to occur.

Assuming a Keplerian orbit, this ratio is equal to ~\citep{freitag02}
\begin{equation}
\frac{E_{\rm orb}}{E_{\rm bind}} = 4.8 (1-e) \frac{GM}{c^2 r_p}
         \left(\frac{R_*}{10{\rm km}}\right)
                 \left(\frac{m}{1.4M_{\odot}}\right)^{-1},
          \label{enrat}
\end{equation}
where we have assumed that the star has zero kinetic energy at infinity.
(Assuming
that the stellar velocity is $10$ km s$^{-1}$ at infinity changes this
result
by only $2.3\times10^{-9}$ for a $1.4 M_{\odot}$ NS of radius
$10$km.)  For an inspiral into a Schwarzschild BH, plunge occurs
when
$c^2 r_p (1+e) = 2 (3+e)GM$; therefore for any eccentricity we have $(1-e)
GM/(c^2 r_p) < 1/6$ at plunge. This means that the energy ratio defined
in~\erf{enrat} can only be greater than $1$ for $R_* > 12.5$~km. Tidal
disruption due to heating is very unlikely to occur. This conclusion also
applies to BHs of moderate spin. For an orbit that is circular at
plunge into a BH with spin $\chi=0.35$, the ratio $E_{\rm
orb}/E_{\rm
bind}$ is approximately equal to $1$ at ISCO for $R_* = 10$km.

If systems existed in which an NS was on a prograde inspiral orbit 
into a rapidly spinning BH, the periapsis at plunge would be much 
closer to the central body and the energy ratio would exceed unity 
at plunge. However, the energy ratio would still be small. The 
radius of the ISCO for a BH of spin $\chi=0.9$ is at $c^2 r_p = 2.32 
GM$, at which radius $E_{\rm orb}/E_{\rm bind}\sim 2$ for $R_* = 
10$km. The disruption criterion that $E_{\rm orb}/E_{\rm bind}\sim 
1$ assumes that the orbital energy is dissipated entirely by tidal 
interactions. In practice, the inspiral will mainly be driven by GW 
emission, since most of the orbital energy is lost in the regime 
where GW emissions are quite significant. Tidal dissipation would 
have to occur on a very short timescale to dominate over 
gravitational radiation reaction effects, and this will not happen 
in practice. We can thus conclude that disruption of the NS due to 
tidal heating will not occur. This is in contrast to main-sequence 
stars that, being less compact, will be disrupted before reaching 
the ISCO \citep{freitag02}. We note that this conclusion does not 
change when the relativistic orbital energy is used in place of the 
Keplerian expression.

\subsubsection{Tidal Capture} \label{tidalcap}

Although tidal interactions should not shorten the inspiral by causing
disruption of the NS, if orbital energy and angular momentum of
the
binary are lost into normal modes of the star, the inspiral trajectory
will be
modified. In principle, this could modify the capture rate and the typical
eccentricities expected at plunge. Significant oscillations are only
likely to be
excited by tidal interactions if the orbital frequency is comparable to
the
frequency of normal modes in the NS. We can estimate the latter
from
the frequency associated with the free-fall time in the NS:
\begin{equation}
\omega_{\rm osc} \approx \frac{\sqrt{2}}{\pi}\,\sqrt{\frac{Gm}{R_*^3}} =
         5.9{\rm kHz} \,
         \left(\frac{m}{1.4M_{\odot}}\right)^{\frac{1}{2}}
         \left(\frac{R_*}{10 {\rm km}}\right)^{-\frac{3}{2}} \label{omosc}.
\end{equation}
This is just an approximation, but it gives the correct order of magnitude
for
the normal mode frequency. \citet{press77} computed normal
modes using a polytropic stellar model with index $n=3$. They found an
$f$-mode
frequency that agrees with Eq.~\erf{omosc}, but with a prefactor of $6.2$
kHz
instead of $5.9$ kHz.

Other stellar modes, in particular $g$-modes, can have significantly lower
frequency and thus will be excited earlier in the inspiral. Press and
Teukolsky tabulate frequencies for $g$-modes up to $g_{14}$, which has a
frequency a factor of $0.15$ smaller than the $f$-mode. An $n=3$ polytrope
is
not a good model for an NS, but it still provides a reasonable
estimate of the frequency range for thermal $g$-modes. NSs also
support crustal $g$-modes that arise from chemical stratification and
core
$g$-modes that arise from stratification in the number densities of
charged
particles. \citet{finn87} computed frequencies of crustal $g$-modes in
zero-temperature NSs, using a range of stellar models. He found
that
the longest period modes had periods of $\sim20$ms. \citet{reisen92}
computed
the frequencies of core $g$-modes, and found that these have similar
frequencies to the crustal modes. Taking $\sim 50$ms as a reasonable
maximum
for the $g$-mode period gives a frequency of $20$Hz.

Inertial ($r$-)modes in rotating NSs typically have frequencies of 
the order of the spin frequency of the NS ($f \sim 10-100{\rm Hz}$). 
\cite{Ho99} examined the excitement of $r$-modes by Newtonian tidal 
driving and found that this was fairly weak. However, \cite{FR07} 
computed the effect of post-Newtonian gravitomagnetic driving and 
found that this was significantly greater. For rapidly rotating NSs, 
the inertial-frame frequency can be much smaller than the 
corotating-frame frequency, which allows $f$- and $p$-modes to be 
excited~\citep{Ho99}. This requires very rapid NS rotation, $f_{\rm 
rot} \sim 500 {\rm Hz}$. \cite{Ho99} examined such modes in the 
context of comparable mass binaries but concluded that such NS 
spins were unlikely to be found in binary systems. In the IMRI case, 
where a free NS is captured, the NS spin could be much higher in 
principle, making these modes potentially interesting. \cite{Ho99} 
and \cite{FR07} considered only modes in the LIGO frequency range, 
$10 {\rm Hz} < f < 1000 {\rm Hz}$, but the mode spectrum extends to 
lower frequencies. However, the frequency at which each resonance 
occurs is a single-valued function of the spin of the NS.

We compare these frequencies to the orbital frequency of a prograde
circular
orbit at radius $r$:
\begin{equation}
\omega_{orb} = 0.65{\rm kHz}\left[\left(\frac{c^2 r}{GM}
         \right)^{\frac{3}{2}} + \chi\right]^{-1}
         \left(\frac{M}{50M_{\odot}}\right)^{-1} \label{omorb}
\end{equation}

Any NS that comes within a distance $\approx 280 GM/c^2$ from an 
$M=100\ M_\odot$ IMBH will be directly captured as a result of GW 
emission. The additional energy lost in tidal interactions could 
increase this capture cross section. However, for $r=300GM/c^2$ 
(cf.~\S~\ref{dircapt}), $\omega_{\rm orb}=0.13$ Hz, which is much 
less than the frequency of oscillations of the star. The $g$-mode 
frequency is $2$ orders of magnitude higher than the orbital 
frequency at that radius and so it is unlikely to be significantly 
excited. The $g$-mode frequencies become comparable to the orbital 
frequency for a Schwarzschild BH when $c^2 r \lesssim 10GM$. Thus, 
$g$-modes are likely to be excited in the late stages of inspiral, 
but not earlier. As mentioned above, the spectrum of NS $r$-modes 
and the $f$- and $p$-mode resonances of rapidly rotating NSs extend 
to low frequencies~\citep{Ho99,FR07}. However, the resonant 
frequencies are determined by the NS spin, so it would require 
extreme fine tuning for a given NS to be captured at precisely the 
periapsis that allows excitement of those modes. The capture rate is 
unlikely to be increased by this mechanism either, although these 
modes could also be excited later in the inspiral.

\citet{press77} provide an expression for the energy
dissipated in tides in an object of mass $m$ and radius $R_*$ that passes
a
point mass of mass $M$ on a Keplerian orbit with periapsis $R_{\rm min}$:
\begin{equation}
\Delta E_{\rm tidal} = \left(\frac{Gm^2}{R_*} \right)
         \left(\frac{M}{m}\right)^2 \left(\frac{R_{*}}{R_{\rm
min}}\right)^6
         T_2\left(\sqrt{\frac{m}{M}}
         \left[\frac{R_{\rm min}}{R_*} \right]^{\frac{3}{2}}\right)
\label{PTenloss}
\end{equation}
This expression is integrated over all thermal normal modes, including
$g$-modes up to $g_{14}$. Once again, this result is based on an $n=3$
polytropic stellar model, which is not a good model of an NS.
However,
it should provide an order-of-magnitude estimate for the energy lost in
thermal
modes. In Eq.~\erf{PTenloss} we include only the $l=2$ modes, since other
modes
are suppressed by $(R_{*}/R_{\rm min})^2 \ll 1$ relative to these modes.
We
also take the extreme mass ratio limit $M \gg m$. The function $T_2(\eta)$
behaves as $T_2(\eta) \sim 0.65 \eta^{-2.34}$ at large $\eta$ (we have
derived
this ``by eye'' from Figure~1 in~\cite{press77}). We can thus compute the
ratio of the energy dissipated in tides to the energy dissipated in GW
emission, $\Delta E_{\rm GW} = [85\pi m^2 /(12\sqrt{2}\,Mc^5)]\,(GM/R_{\rm
min})^{7/2}$, for an object on a parabolic Keplerian orbit with periapsis
$R_{\rm min}$:
\begin{equation}
\frac{\Delta E_{\rm tidal}}{\Delta E_{\rm GW}} \approx 0.05
         \left(\frac{GM}{c^2\,R_{\rm min}}\right)^{6.01}
         \left(\frac{R_*}{20{\rm km}}\right)^{8.51}
         \left(\frac{M}{50\Msun}\right)^{-5.34}
         \left(\frac{m}{1.4\Msun}\right)^{-3.17}
\label{PTrat}
\end{equation}
It is clear that, under these model assumptions, the tidal perturbation to
the orbit at capture is always much weaker than the perturbation induced by
GW emission. For comparison, since $\Delta E_{\rm GW} \propto r_p^{-7/2}$, a $10\%$ increase in the energy lost in a single pass by the central BH increases the minimum periapsis required for direct capture by only a factor of $1.1^{2/7} \approx 1.03$ or $\sim 3\%$.

The above arguments indicate that the excitement of NS modes will 
neither increase the capture rate nor lead to NS disruption during an 
IMRI. However, orbital energy lost into oscillations could modify 
the inspiral trajectory by either causing a cumulative phase shift 
in the emitted GWs or changing the eccentricity of the orbit in 
the LIGO band. \cite{FR07} calculated the phase difference in the 
GWs that arises from the excitement of $r$-modes, finding $\Delta 
\Phi \sim 3.4 R_{10}^4 f_{s100}^{2/3}M_{1.4}^{-1} m_{1.4}^{-2} 
(M_{1.4}+m_{1.4})^{-1/3}$, where $R_{10}$ is the NS radius in units 
of 10km, $M_{1.4}$/$m_{1.4}$ are the masses of the primary/secondary 
in units of $1.4M_{\odot}$, and $f_{s100}$ is the spin frequency (or 
$r$-mode frequency) in units of $100$ Hz. For an IMRI with 
$M=50M_{\odot}$, this gives $\Delta\Phi \sim 0.003$ if we set 
$R_{10} = f_{s100} = m_{1.4} = 1$. Typically we require a phase 
shift of $\Delta\Phi \sim 1$ for an effect to be observable, so the 
excitement of $r$-modes will not leave an imprint on the inspiral. 
The phase shift induced by the resonant excitement of $f$- and 
$p$-modes in rapidly rotating NSs can be significantly higher. 
\cite{Ho99} quote $\Delta\Phi \sim 234 m_{1.4}^{-4.5} R_{10}^{3.5} 
m_{1.4}^2/(M_{1.4}(m_{1.4}+M_{1.4})) f_{gw100}^{-1}$ for the most 
extreme case of the $(22, 2)$ $f$-mode resonance (with the same 
notation as before but now denoting the GW frequency in units of 
$100$ Hz by $f_{gw100}$). For an $M=50M_{\odot}$ IMRI, this gives 
$\Delta\Phi \sim 0.2 f_{gw100}^{-1}$. This could be a measurable 
shift if the resonance is excited near $10$ Hz. However, the phase 
shift is only this large for IMBHs at the low-mass end of the IMRI 
range, and provided that the NS spin is tuned to ensure that the 
resonance is excited near $10$ Hz. More work will be needed to 
quantify how large a phase shift would be measurable with Advanced 
LIGO, accounting for correlations between the waveform parameters.

\begin{figure}
\plotone{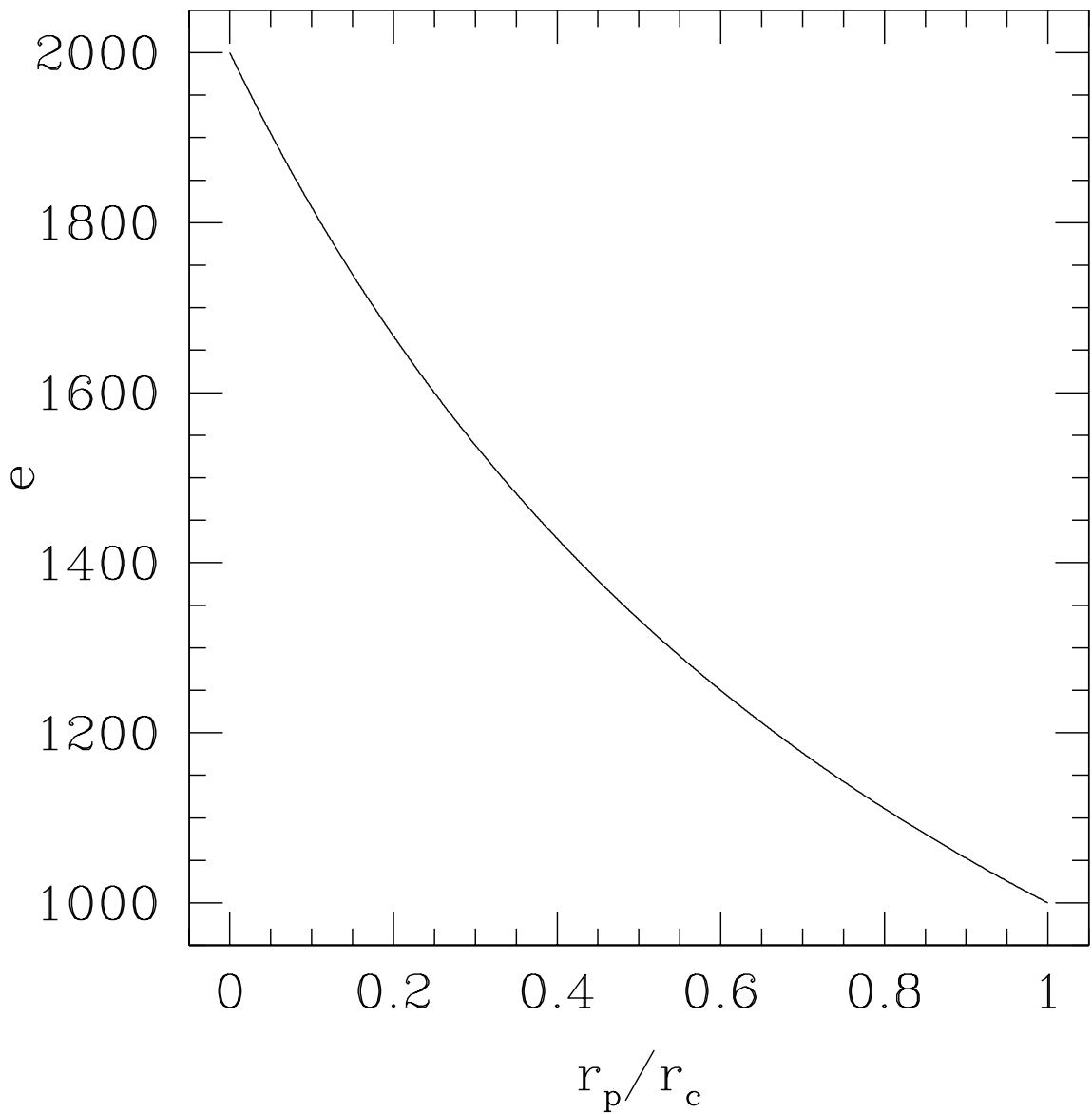}
\figcaption{Tidal-dissipation-driven inspiral in phase space for an
inspiraling
star with initial eccentricity of $e=1$ and initial periapsis $r_p =
1000\,r_c$.
The plot shows eccentricity on the horizontal axis and the ratio $r_p/r_c$
on the vertical axis. The radius $r_c$ characterizes the frequency of
normal
modes in the star as defined by Eq.~\erf{omzz}.\label{tidalinffig}}
\end{figure}

We can estimate qualitatively what effect tidal dissipation would have on
the orbital eccentricity and periapsis. The phase-space trajectory that an
inspiral follows is determined entirely by the ratio $\rmd E/\rmd L_z$.
Assuming a Keplerian orbit, we have
\begin{equation}
\frac{\rmd r_p}{\rmd e} = \frac{r_p\left(2\,\sqrt{(1+e)GM} -
r_p^{\frac{3}{2}}
          \,\rmd E/\rmd L_z\right)}
         {(1+e)\,r_p^{\frac{3}{2}}\,\rmd E/\rmd L_z
         -2(1-e)\sqrt{(1+e)GM}}. \label{eveq}
\end{equation}
We now suppose that the inspiral was driven entirely by tidal dissipation.
Typically the dominant excited mode would be an $m=2$ mode, for which
$\Delta
L_z = 2 \Delta E/\omega_{00}$, where $\omega_{00}$ is the frequency of the
mode
(this assumes that the stellar oscillations can be modeled as a linear
Lagrangian system; \citealt{friedman78}). We write
\bel{omzz}
\omega_{00} = \sqrt{\frac{GM}{r_c^3}},
\ee
where $r_c$ is the radius of the circular (Keplerian) orbit that would
have
the same frequency as the $m=2$ mode. With this substitution, equation
\erf{eveq} defines the evolution of $r_p/r_c$ over the inspiral. Equations
\erf{omosc} and \erf{omorb} indicate that the capture periapsis, $r_p^0$,
will
typically be much  greater than $r_c$. Solutions with $r_p^0 > 2^{5/3}
r_c$ are
all qualitatively the same, and we show a typical example in
Figure~\ref{tidalinffig}, for capture periapsis of $1000\,r_c$ and a
capture
eccentricity of $1$. For a $100\Msun$ IMBH, taking $\omega_{00} = 6$ kHz
yields
$c^2\,r_c \approx 0.5 GM$, so this figure represents a capture at $r_p
\approx
500\ GM/c^2$, the upper end of the allowed direct capture range for an
$m=10\
M_\odot$ BH. The figure shows the inspiral in
eccentricity-periapsis
space. Under this simple model of tidal interactions, the periapsis
increases
while the eccentricity decreases. In practice, the inspiral will be driven
by a
combination of GW emission and any tidal dissipation that
occurs. These results suggest that tidal effects would tend to make the
eccentricities at plunge smaller than they would be for inspirals driven
by
radiation reaction alone.


Equations \erf{omosc}--\erf{PTrat} indicate that normal modes are 
unlikely to be excited during an inspiral into an IMBH, although 
high-order $g$-modes, $r$-modes, and $f$-modes in rapidly rotating 
NSs might be excited during the very late stages of inspiral. Thus, 
we can safely ignore the effect of tides on the capture rates. Tidal 
effects could modify the inspiral, although the above calculation 
indicates that this should not modify our conclusions about the 
typical eccentricities at plunge. The excitement of $f$-modes might 
leave a measurable imprint on the GW signal. However, the induced 
phase shift is only marginally detectable, and this mechanism 
requires the NS to be rapidly rotating.

\section{Event Rates}\label{sec:rates}

In this section we estimate the rate of IMRIs in globular clusters 
detectable by Advanced LIGO. To do this, we must consider three 
elements: (1) the distance sensitivity of the detectors to GWs from 
IMRIs (and hence the volume of the universe the detectors can see), 
(2) the number density of globular clusters, and (3) the rate of 
IMRIs per globular cluster.

\subsection{Advanced LIGO IMRI Sensitivity}
\label{LIGOrange}

For GW sources with known waveforms (or at least waveforms
well
approximated by analytic or numerical techniques), matched filtering is
used to
search for signals in GW detector
data~\citep{Wainstein:1962,Allen:2005fk}. The S/N $\rho$
of a template $h(t)$ in data $s(t)$ collected by a detector that has
one-sided noise power spectral density $S_n(|f|)$ is given by
\begin{equation}
\rho = \frac{4}{\sigma}
\int_0^\infty \frac{|\tilde{s}(f) \tilde{h}^\ast(f)|}{S_n(|f|)}\, df,
\label{eq:snr}
\end{equation}
where $\tilde{s}(f)$ is the Fourier transform of the signal $s(t)$,
$\tilde{h}(f)$ is the Fourier transform of the inspiral template $h(t)$,
the asterisk denotes complex conjugation, and $\sigma$ is defined by
\begin{equation}
\sigma^2 = 4 \int_0^\infty \frac{|\tilde{h}(f)|^2}{S_n(|f|)}\, df.
\label{snrsq}
\end{equation}
This definition of S/N follows the normalization of 
\cite{Cutler:1994ys} and~\cite{Allen:2005fk}. We place the template 
$h(t)$ at a canonical source distance of $1$~Mpc and choose the 
optimal orientation of the detector to maximize the S/N, and so the 
maximum distance to which a single detector matched filter search is 
sensitive at a given S/N $\rho$ is given by $D = 
{\sigma}/{\rho}$~Mpc.  (This is the same quantity as the ``inspiral 
horizon distance'' used by the LIGO and Virgo 
Collaborations~\cite{S3S4}.)

To compute the sensitivity of a single Advanced LIGO detector to IMRIs, we
need
to compute the quantity $\sigma^2$ defined in Eq.~\erf{snrsq} using a
particular waveform model. We have done this with waveforms based on BH
perturbation theory~\citep{finnthorne00}, which are valid in the limit
$m/M \ll
1$. The waveforms, which include non-quadrupolar harmonics of the orbital
frequency in addition to the dominant quadrupolar harmonic, are described
in
Appendix~\ref{harmapp}, where we also discuss the relative S/N contributed
by
the four lowest harmonics. The noise power spectral density $S_n(|f|)$ was
taken from~\cite{Fritschel}.  GW detectors have an
orientation-dependent response.  The relation between the {\it range} $R$
(defined as the radius of a sphere whose volume is equal to the volume of
the
universe in which inspiral sources could be detected with an S/N threshold
of
$\rho$) and maximum distance $D$ at a fixed S/N is given by $R = D /
2.26$~\citep{Finn:1992xs}.

We assume a value of $\rho = 8$ for the threshold S/N required for a 
detection, since this is the value typically used to compute Initial 
LIGO detection ranges for comparable-mass black hole 
binaries~\citep{S3S4}. This is a reasonable approximation, as a 
binary black hole inspiral with a total mass of $6\,M_\odot$ has 
approximately $500$ GW cycles between the $40$~Hz low-frequency 
cutoff of Initial LIGO and coalescence --- roughly the same number 
of GW cycles that an IMRI signal in Advanced LIGO will have between 
the Advanced LIGO low-frequency cutoff of $10$~Hz and coalescence. 
The threshold will be computed more accurately when an IMRI search 
is implemented and the amount of non-stationarity of the Advanced 
LIGO data is known. If the $\rho=8$ threshold cannot be achieved in 
practice (or if it can be improved), then the detection rates 
derived below can be scaled appropriately.

Advanced LIGO will consist of a network of three 4-km detectors. Demanding
that
GWs are found coincident in all three detectors increases
the
network range by a factor of $\sqrt{3}$ relative to the range of a single
detector at a given S/N (due to the lower false alarm rate of the
network).
Fig.~\ref{Fig-range} shows the range $R$ of a network of three Advanced
LIGO
detectors for circular-equatorial-orbit IMRIs of $m=1.4\Msun$ objects into
a
Kerr IMBH of mass $M$, assuming that the network S/N required for a
confident
detection was $\rho=8$. This is equivalent to the range of a single
detector
with S/N of $\rho = 8/\sqrt{3}$.  The $\chi=0$ (non-spinning IMBH) line
in
Fig.~\ref{Fig-range} is well-approximated by a quadratic fit:
\begin{equation}
R \approx \sqrt{m/M_\odot} \times
         \left[800 - 540\left(\frac{M}{100\ M_\odot}\right)
         +  107\left(\frac{M}{100\ M_\odot}\right)^2\right] \,\textrm{Mpc}.
\label{eq:rangefit} \end{equation}


The scaling of the range in Eq.~(\ref{eq:rangefit}) as $\sqrt{m}$ does not
follow
from the fit, but rather from the following reasoning.  The amplitude of
GWs from IMRIs will scale linearly  with the mass of the
smaller object $m$, but the number of cycles in the LIGO band will also
drop by
roughly a factor of $m$.  Hence, the total S/N will
grow as $\sqrt{m}$, so inspirals of more massive COs will be
seen a factor of $\sqrt{m}$ farther away.

The combination of the spin of the central object and the 
inclination of the orbital plane of the inspiraling particle will 
have a significant effect on the signal from an IMRI.  The frequency 
of the ISCO is much higher for prograde inspirals into rapidly 
spinning BHs than for inspirals into non-spinning holes; the S/N can 
be strongly enhanced for such orbits. Conversely, retrograde 
inspirals will have lower S/R.  Averaging over random inclination 
angles, \citet{Mandel} computed the ratio between (1) the detection 
range for Advanced LIGO in a universe uniformly populated by IMBHs 
of a given mass and spin and (2) the detection range in a universe 
with an equal density of Schwarzschild IMBHs with the same mass.  He 
found that the detection range can be enhanced by a factor of $1.7$ 
($3.8$) for maximally spinning Kerr BHs with $M=100\Msun$ 
($M=200\Msun$); the increase in the volume of observable space and, 
hence, the event rates, is the cube of these numbers.

If IMBHs grow mainly by random mergers, they will not be rapidly spinning
as
the contributions of subsequent mergers to the hole's spin largely cancel
out.
The angular momentum imparted to the IMBH by a CO is $L_{\rm
obj}
\propto m M$, since the radius at ISCO is $r_{\rm ISCO} \propto M$.  This
causes
the dimensionless spin parameter of the hole $\chi=S_1/M^2$ to change by
$\sim
L_{\rm obj}/M^2 \propto m/M$.  After $\sim M/m$ such mergers, necessary
for the
hole to grow to mass $M$, the typical spin of the hole will be $\chi \sim
\sqrt{m/M}$.  More precise calculations \citep{HB03, M02, Mandel} show
that the
spin of IMBHs involved in LIGO IMRIs will rarely exceed $\chi=0.3$ for
IMBHs
that gained a significant fraction of their mass via minor mergers.  For
small
values of $\chi$, Eq.~(24) of \cite{Mandel} yields a correction to the
range
presented in Eq.~(\ref{eq:rangefit}) due to the inclusion of the IMBH
spin;
the detection range in Mpc as a function of $M$, $m$, and $\chi$ is
\begin{equation}\label{range}
\frac{R}{\rm Mpc} \approx \left[1 + 0.6\ \chi^2
         \left(\frac{M}{100\ M_\odot}\right)\right]
         \sqrt{\frac{m}{M_\odot}}
         \left[800 - 540
         \left(\frac{M}{100\ M_\odot}\right) +
         107\left(\frac{M}{100\ M_\odot}\right)^2\right].
\end{equation}

This range estimate does not include the cosmological redshift.  The
redshift
due to the expansion of the universe decreases the frequency of the
GWs.  For $M \sim 100\ M_\odot$ IMRIs, the redshifted GWs
will
lie in a less sensitive part of the LIGO noise curve, thereby reducing the
range. For IMRIs detectable with Advanced LIGO, redshifts are typically
$\lesssim 0.2$; for example, the inspiral of a $1.4 M_\odot$ NS
into a non-spinning $100 M_\odot$ IMBH is visible to a redshift of $0.09$.
We
estimate that for typical sources, properly including the redshift reduces
the
Advanced LIGO event rate by $\sim 10\%$.

Advanced LIGO will have several parameters that may be tuned during 
the operation of the detector to optimize the noise power spectral 
density (PSD) in order to search for specific sources.  These 
tunable parameters include the laser power and the detuning phase of 
the signal recycling mirror.  If a noise PSD optimized for 
detections of CO--IMBH binaries is used instead of the default PSD 
assumed in Fig.~\ref{Fig-range}, the range for such sources is 
increased by a factor of $\sim 1.5$, corresponding to an event rate 
increase by a factor of $\sim 3.5$.

\subsection{Number Density of Globulars with a Suitable IMBH}
\label{IMBHdensity}

The second element in the rate calculation is the number density of 
globular clusters that have an IMBH in the relevant mass range.  
This is highly uncertain. To contribute significantly, a cluster 
must have had a sufficiently small initial relaxation time to allow 
the formation of an IMBH through some mild runaway process when the 
cluster was young, yet not have formed an IMBH with $M>350\,M_\odot$ 
(since this would put IMRIs beyond the Advanced LIGO frequency 
range).  Recent theoretical arguments by Trenti and colleagues 
\citep{Heggie06,Trenti06a,Trenti07,Trenti06b} suggest that 
dynamically old globulars with large core to half-mass radius ratios 
have been heated by a $\sim 1000\,M_\odot$ IMBH, so these clusters 
would not contribute to the Advanced LIGO IMRI rate.  Note, however, 
that~\cite{Hurley07} has shown that current observations of the 
core-to-half-light ratios in globulars do not require 
$1000\,M_\odot$ BHs in most clusters. Core-collapsed globular 
clusters, which constitute $\sim$20\% of all globular 
clusters~\citep{Phinney91}, may contain IMBHs of the right mass. We 
will parametrize the unknown fraction of relevant globular clusters 
by some fraction $f$.  Globular clusters have a space density of 
$8.4\ h^3\ {\rm Mpc}^{-3}$ \citep{PZ}, which for $h=0.7$ yields 
$2.9\ {\rm Mpc}^{-3}$.  Therefore, we will use the number density 
$\sim 0.3\ (f/0.1)\ {\rm Mpc}^{-3}$. This factor $f$ depends on both 
the number of clusters with an IMRI in the right mass range and the 
number of clusters that have been in a state of core-collapse long 
enough for the binary hardening mechanism to occur. These factors 
are degenerate, however, since clusters with heavier IMBHs will not 
be in a state of core collapse, as described above. The fraction $f$ 
also depends on what proportion of the objects merging with the IMBH 
are COs as opposed to main-sequence stars. Our Monte Carlo 
simulations, which were discussed earlier, indicate that this 
proportion is close to 1.

The fraction $f$ of globular clusters containing IMBHs may be 
further lowered by ejections of IMBHs from their clusters by recoil 
kicks imparted to the IMBHs by dynamical processes and by 
gravitational radiation emission.  If the kick exceeds $\approx 50\ 
{\rm km}\ {\rm s}^{-1}$, which is the escape velocity from a massive 
globular cluster, the IMBH will escape from the cluster, thereby 
becoming unavailable for future events. Kicks can arise from the 
process of hardening via three-body encounters~\citep{KHM93, SH93, 
GMH04, GMH06}. G\"ultekin, Miller and Hamilton (2006) show (cf. 
their Fig.~12) that when the seed mass is $100 \Msun$, only about 
50\% of all BHs grow to $300 \Msun$ without being ejected, and this 
fraction drops to $10\%$ for a seed mass of $50 \Msun$.

Kicks also arise from GW emission. During the last stages of the 
merger of unequal mass BHs, a net flux of angular momentum will be 
carried away by the GWs, imparting a kick to the resulting 
BH~\citep{Per62, Bek73, Fit83, FD84, RR89, W92, FHH04, BQW05, DG06, 
Herrmann06, Baker06, Gonzalez06, Sop06}. The most recent results on 
merger velocity kicks, based on numerical relativity, show that the 
kick velocity for a non-spinning central object depends on the 
symmetric mass ratio $\eta=mM/(m+M)^2$ as $V_{\rm kick} \approx 
12000 \eta^2 \sqrt{1-4\eta} (1-0.93\eta)$~km 
s$^{-1}$~\citep{Gonzalez06}.  The requirement $V_{\rm kick} < 50$ km 
s$^{-1}$ places an upper limit on $m$ of $q=m/M \lesssim 0.08$.

If the IMBH is rapidly spinning, recent numerical relativity results 
suggest that the kick can be a lot higher~\citep{Baker07, 
Campanelli07a, Campanelli07b, Gonzalez07, Herrmann07, Koppitz07}. 
\cite{Baker07} and \cite{Campanelli07a} provide a fit to numerical 
relativity results that gives the kick as a function of the various 
orbital parameters.  This formula indicates that if the IMBH has 
moderate spin $\chi \lesssim 0.5$ and the secondary is non-spinning, 
then we require $q \lesssim 0.05$ to ensure that the IMBH has a high 
probability of remaining in the globular cluster today after 
undergoing multiple mergers. This constraint can be relaxed to $q 
\lesssim 0.067$ if $\chi \lesssim 0.3$. If the objects merging with 
the IMBH are BHs with a mass of $10\Msun$, this constrains the 
initial IMBH mass to be $M \gtrsim 150\Msun$. If the merging objects 
are $1.4\Msun$ NSs, even IMBHs with a seed mass of $50\Msun$ are 
safe from ejection.

As argued earlier, mergers with BHs are likely to be important early 
in the IMBH evolution, when its mass is smaller, with NS mergers 
becoming dominant later. This could mean that a significant number 
of IMBHs were ejected from globular clusters early in their 
evolution. However, without firm knowledge of the initial seed 
masses of IMBHs or the relative number of mergers with BHs and NSs 
that each IMBH undergoes, it is impossible to draw definitive 
conclusions. We normalize $f$ to $10\%$ in the rate calculations 
that follow, but we emphasize that this quantity is highly uncertain 
at present.

\subsection{IMRI Rate per Globular Cluster and Event Rate}

The final contribution to the rate estimate is the merger rate per 
globular cluster. Existing numerical simulations of globular 
clusters suggest that mergers in the subcluster of $\sim 10 M_\odot$ 
BHs at the center of the globular cluster can lead to the creation 
of IMBHs with masses up to $\sim 350 M_\odot$ in $\sim 10^{10}$ yr 
\citep{OL06}.  However, the results of such simulations are very 
sensitive to the choice of cluster models and to assumptions about 
kick velocities, the interaction between the BH subcluster and the 
rest of the cluster, etc. Therefore, we present two methods for 
computing the rate per globular: (1) an upper limit independent of 
cluster model and (2) an estimate based on a more realistic model 
for cluster dynamics.

We estimate a theoretical upper limit on the IMRI event rate in a 
globular cluster using the following method, originally suggested by 
\citet{Phinney}.  We assume that each globular cluster has a BH that 
grows from $M \sim 50 \Msun$ to $M\sim 350 \Msun$ by capturing a 
sequence of COs of identical mass $m$ over the age of the cluster. 
Then $300 \Msun/m$ captures will happen in each globular cluster in 
$\sim 10^{10}$ yr. This leads to a rate of $(300\ \Msun)/m \times 
(10^{10}\ {\rm yr})^{-1}$ per cluster.

Although this rate is plausible, it may be a significant 
over-estimate for several reasons. First, it assumes that all the 
mass that the IMBH acquires in growing from $M \sim 50 \Msun$ to 
$M\sim 350 \Msun$ comes from mergers with COs. In practice, the IMBH 
will also acquire mass via gas accretion, and by captures of 
main-sequence stars and white dwarfs, which will be tidally 
disrupted before becoming significant GW sources but will still add 
mass to the IMBH.  Second, this estimate does not include the 
likelihood that the merger product will be kicked out of the cluster 
through recoil, as discussed in the previous section.  Third, this 
estimate assumes that the rate at which the IMBH grows via IMRIs 
from $50\ \Msun$ to $350\ \Msun$ is constant in time.  However, 
Advanced LIGO can only detect mergers that occurred at distances 
$\lesssim 1$ Gpc, i.e., relatively recently, so the relevant rate is 
the rate late in the history of the globular cluster, which is 
likely to be much lower. For example, \cite{OL06} found in their 
numerical simulations that the rate dropped from $\sim 10^{-7}$
to $\sim 3\times 10^{-10}$ yr$^{-1}$ after $10^{10}$ yr for some 
plausible cluster models.

For the theoretical upper limit, the total rate is given by $\alpha 
\overline{V(M,m,\chi)}$, where $\alpha \sim 0.3\ (f/0.1)\ {\rm 
Mpc}^{-3}\ (300\ \Msun)/m\ (10^{10}\ {\rm yr})^{-1}$ is the IMRI 
rate in the universe, $V(M,m,\chi)=(4/3)\pi R^3$ is the volume in 
which Advanced LIGO can see an event, and on overbar, 
$\overline{V}$, denotes the average over mass $M$ in the range 
between $50 \Msun$ and $350 \Msun$.  If we take $f=0.1$, $\chi=0.2$ 
as the typical IMBH spin, and all inspiraling objects are $1.4 \Msun$ 
NSs, the event rate is $\approx 3$ yr$^{-1}$; 
if $f=0.1$, $\chi=0.2$, 
and inspiraling objects are $10 \Msun$ BHs, the event rate is 
$\approx 10$ yr$^{-1}$.  These values are based on the range fit in 
Eq.~(\ref{range}), so they assume that orbital frequency harmonics 
through $m=4$ are included in the data analysis, but cosmological 
redshift and Advanced LIGO optimization are not included.  When all 
of these considerations are taken into account, a theoretical 
upper-limit estimate suggests that Advanced LIGO may detect up to 
$30$ IMRIs per year.  A similar estimate for Initial LIGO shows 
that because of lower overall sensitivity and a higher low-frequency 
cutoff ($40$ Hz for Initial LIGO vs.~$10$ Hz for Advanced LIGO), the 
upper limit on the Initial LIGO IMRI rate is only about $1/1000$ 
events yr$^{-1}$.

A more realistic estimate is based on the assumption that the 
hardening of a CO--IMBH binary via three-body interactions 
represents the primary capture mechanism leading to IMRIs.  The rate 
for IMRIs created by this scenario is $\approx 3 \times 
10^{-9}$~yr$^{-1}$ per globular cluster for NS--IMBH IMRIs and 
$\approx 5 \times 10^{-9}$~yr$^{-1}$ for BH--IMBH IMRIs [see 
\S~\ref{harden}]. Hence, the NS--IMBH IMRI rate in the local 
universe is $\alpha\approx 10^{-9}\ (f/0.1)\ {\rm Mpc}^{-3}\ {\rm 
yr}^{-1}$, while the BH--IMBH IMRI rate is $\alpha\approx 1.5 \times 
10^{-9}\ (f/0.1)\ {\rm Mpc}^{-3}\ {\rm yr}^{-1}$.  If we assume that 
all IMBHs have a mass $\sim 100 \Msun$ and $f=0.1$, this yields an 
Advanced LIGO rate of one IMRI per $3$ years if the typical CO is an 
NS or $10$ IMRIs per year if the typical CO is an $m=10\ M_\odot$ 
BH.  If the interferometer is optimized for the detection of IMRIs, 
the NS--IMBH and BH--IMBH rates are increased to $1$ and $30$ events 
yr$^{-1}$, respectively.

In addition to detections of inspirals, Advanced LIGO could also 
detect the ringdown of an IMBH following a merger.  This possibility 
is discussed in Appendix~\ref{ringdownapp}.

\section{Effect of Eccentricity on Matched Filter Searches}\label{sec:ecc}

As discussed in \S~\ref{LIGOrange}, matched filtering is used to 
search for GWs with known waveforms in detector noise. In order to 
be an optimal search technique, the matched filter requires accurate 
templates that correctly model the signals being 
sought~\citep{Wainstein:1962}. Since source parameters (e.g., the 
masses and the IMBH spin) can vary, the matched filter is 
constructed for a ``bank'' of templates: a set of waveform models 
that depend on the parameters that characterize the source. The 
accuracy of a template bank is characterized by the fitting factor 
(FF)~\citep{Apostolatos}, which measures the overlap between the GW 
signal and the nearest template. A fitting factor close to unity 
indicates that the templates are accurate for detection of the 
desired signals. A fitting factor less than unity will mean that we 
are unable to detect a fraction $(1 - \mathrm{FF}^3)$ of the 
theoretically detectable events. (The quantity $1 - \mathrm{FF}$ is 
often referred to as the mismatch.) To search for signals, template 
banks are constructed so that the mismatch between any desired 
signal and the nearest template does not cause an unacceptable loss 
in S/N (typically $\mathrm{FF} \approx 0.97$ for LIGO).

In this section we examine the effect of eccentricity on searching 
for IMRI signals in Advanced LIGO detectors. The effect of 
eccentricity on the fitting factor was previously examined 
by~\cite{Martel99}, and it was found that the fitting factor between 
a circular and eccentric waveform template was high provided that $e 
\lesssim 0.2$. However, their results do not apply directly to IMRIs 
since they computed fitting factors only for binaries with mass 
ratios close to $1$, and used the first-generation LIGO noise curve.

We consider a matched-filter search for IMRIs and determine the loss 
in S/N (and hence range) if eccentricity is not included in the 
template bank; i.e., circular templates are used to search for 
potentially eccentric waveforms.  We compute the fitting factor as 
follows. The template $h(t)$ appearing in the expression for the 
matched filter S/N $\rho$ [Eq.~(\ref{eq:snr})] depends on a number 
of parameters characterizing the source, such as the masses of the 
binary and the time of arrival of the signal. We denote these 
parameters $\vec{\lambda}$ and define the ambiguity function 
$\mathcal{A}(\vec{\lambda})$ by 
\begin{equation} 
\mathcal{A}(\vec{\lambda}) =
         \frac{\langle s | h(\vec{\lambda}) \rangle}
         {\sqrt{\langle s | s \rangle
         \langle h(\vec{\lambda}) | h(\vec{\lambda}) \rangle}},
\end{equation}
where $\langle a | b \rangle$ is the matched filter inner product given by
\begin{equation}
\langle a | b \rangle = 4 \int_0^\infty
\frac{\tilde{a}(f)\tilde{b}^\ast(f)}
         {S_n(|f|)}\, df. \label{eq:innerprod}
\end{equation}

We separate the parameters $\vec{\lambda}$ into $\vec{\lambda} = 
(t_0, \phi_0, \vec{\theta})$, where $t_0$ and $\phi_0$ are the time of 
arrival and phase of the binary, respectively. In the case of 
circular equatorial binaries, it is trivial to maximize over the 
parameters $t_0$ and $\phi_0$ analytically (the phase by projecting 
the signal onto two orthogonal basis vectors and the time by a 
Fourier transform), and so these are called ``extrinsic 
parameters.'' The remaining template parameters $\vec{\theta}$, which 
include the binary masses, eccentricity, and IMBH spin, determine the 
shape of the waveform and are known as ``intrinsic parameters.'' 
For circular inspiral templates, the ambiguity function 
$\mathcal{A}$ reduces to the overlap $\mathcal{O}$, given by
\begin{equation}
         \mathcal{O}(\vec{\theta}) = \max_{t_0, \phi_0}
         \frac{\langle s | h(\vec{\theta}) \rangle}
         {\sqrt{\langle s | s \rangle
         \langle h(\vec{\theta}) | h(\vec{\theta}) \rangle}},
\end{equation}
The fitting factor is given by the maximum of the overlap function over
the
remaining parameters
\begin{equation}
         \mathrm{FF} = \max_{\vec{\theta}} \mathcal{O}(\vec{\theta}).
\end{equation}

For the signal $s(t)$ and template $h(t)$ we use numerical kludge 
waveforms. This is a family of waveforms that were constructed as models 
for extreme mass ratio inspiral systems, in which $m/M \ll 1$. The 
waveform family is constructed by first computing an accurate 
phase-space trajectory by integrating prescriptions for the evolution of 
the orbital elements (the orbital energy, angular momentum, and Carter 
constant, or equivalently the orbital radius, eccentricity, and 
inclination)~\citep{GG06}. The orbit of the small body is then 
calculated by integration of the Kerr geodesic equations along the 
sequence of geodesics defined by the phase-space trajectory. Finally, a 
kludge waveform is generated from the orbit by applying weak-field 
emission formulae~\citep{babak06}. This waveform family predicts the 
inspiral rates for nearly-circular orbits very well~\citep{GG06} and has 
been shown to be extremely faithful (overlaps in excess of $\sim 95\%$ 
over much of the parameter space) to more accurate perturbative 
waveforms~\citep{babak06}. Although the mass ratio of an IMRI system is 
probably too high to make these waveforms accurate as search templates, 
they should provide reliable predictions of the fitting factor.

For these calculations we used $M=100\ \Msun$ for the IMBH mass, $m=1.4\ 
\Msun$ for the companion mass, and considered two spin values $\chi=0$ 
and $\chi=0.2$. We used the Advanced LIGO power spectral density 
$S_n(|f|)$ given by~\cite{Fritschel}.  As discussed above, to compute 
the fitting factor one must maximize over the parameters $\vec{\theta}$ 
of the template. However, we find that even without maximizing over the 
intrinsic parameters, the overlap (and hence the fitting factor) between 
circular and eccentric templates is greater than $0.99$ for 
eccentricities $e < 0.01$, i.e., for more than two-thirds of IMRIs 
formed by direct capture (the mechanism likely to give the largest 
eccentricities). Since we expect that most of the IMRI systems will have 
eccentricities significantly less than $e=0.01$ by the time they have 
entered the Advanced LIGO band, eccentricity will be negligible for data 
analysis and circular templates may be used to search for these systems.

Fig.~\ref{f:overlap} shows the overlap between eccentric signals and 
circular templates for prograde equatorial inspirals and eccentricities 
greater than $0.01$.  Analysis of inclined inspirals demonstrates that 
the overlaps between eccentric signals and circular templates remain 
greater than $0.99$ for eccentricities $e<0.01$ and greater than $0.93$ 
for eccentricities $e<0.05$. Although the overlap decreases for 
eccentricities greater than $0.01$, we anticipate higher values of the 
fitting factor when we maximize over the other intrinsic parameters. An 
interesting question will be to determine whether eccentricities greater 
than $0.01$ can be measured (and thus be used to investigate the 
relative prevalence of the various capture mechanisms) or if 
eccentricity is degenerate with masses and the other intrinsic 
parameters.


\section{Summary}\label{sec:summary}

In this paper we have discussed a potential source of GWs for 
ground-based interferometers: the intermediate mass ratio inspiral 
of a stellar mass CO (an NS or BH) into an IMBH in the center of a 
globular cluster. For IMBHs with masses in the range 
$50-350M_{\odot}$, the GWs emitted will be at frequencies in the 
Advanced LIGO band. We have shown that Advanced LIGO should be able 
to detect the inspiral of a $1.4M_{\odot}$ NS into an IMBH at 
distances up to $700$ Mpc, depending on the mass and spin of the 
IMBH. Assuming that all IMBHs were grown by CO--IMBH mergers gives an 
upper limit on the Advanced LIGO event rate of $\sim 10$ yr$^{-1}$.
We have shown that if the inspiraling CO is an NS, a more likely 
estimate of the rate is one event per $3$ years, while the rate 
for BH--IMBH IMRIs could reach the upper limit. If Advanced LIGO is 
optimized for detections at low frequencies, the event rate 
estimates would increase by a factor of $\sim 3.5$.

We have also discussed four mechanisms by which such IMRI systems 
could form: (1) binary hardening via three-body interactions, (2) 
hardening via Kozai resonance, (3) direct capture, and (4) tidal 
capture of a main-sequence star. In all four cases, we find that the 
residual eccentricity when the inspiral enters the LIGO sensitivity 
band will be small.  Finally, we have estimated the sensitivity of 
Advanced LIGO to the eccentricity of IMRI systems. We have found 
that the eccentricities we expect are negligible for data analysis, 
and therefore circular-orbit templates may be used to search for 
IMRI binaries in Advanced LIGO.

IMRIs are a somewhat speculative source of GWs, since evidence for 
the existence of IMBHs is not yet conclusive. The body of evidence 
is steadily growing, however. Since little is known about the 
abundance of IMBHs in the universe, the event rates presented here 
are naturally somewhat uncertain. However, our results are 
sufficiently promising to make IMRIs a source worth searching for in 
Advanced LIGO data. If IMRI events are detected with Advanced LIGO, 
these will provide irrefutable evidence for the existence of BHs 
with intermediate mass and will provide information on the mass and 
spin of IMBHs, plus the eccentricities of the inspiraling objects. 
This information will be very useful for constraining models of IMBH 
formation and growth and for exploring stellar dynamics in the 
centers of globular clusters.

\acknowledgements
The authors are grateful to Sterl Phinney, Kip Thorne, Yuri Levin, 
Clovis Hopman, and Teviet Creighton for helpful discussions.  I.M.~thanks 
the Brinson Foundation, NASA grant NNG04GK98G, and NSF grants PHY-0601459 
and PHY-0653321 for financial support.  M.C.M.~acknowledges support from 
the National Science Foundation under grant AST 06-07428. J.G.'s work was 
supported by St.~Catharine's College, Cambridge. D.B.~acknowledges support 
from the LIGO Laboratory and the National Science Foundation under grant 
PHY-0601459. LIGO was constructed by the California Institute of 
Technology and Massachusetts Institute of Technology with funding from 
the National Science Foundation and operates under cooperative agreement 
PHY-0107417. This paper has LIGO Document Number LIGO-P070014-00-Z.


\appendix
\begin{center}
{\bf APPENDIX}
\end{center}
\section{Waveforms and Signal-to-Noise Ratio Calculation}
\label{harmapp}

To compute the range to which a source can be seen, as presented in 
\S~\ref{LIGOrange}, we must evaluate the S/Ns of typical sources. To 
do this requires a model of the waveform. In the weak field, 
waveforms may be well approximated by post-Newtonian results. The 
leading-order post-Newtonian result takes the system to be a 
Keplerian binary and estimates the gravitational radiation from the 
leading-order quadrupole formula~\citep{Peters:1963ux,P64}. This 
predicts $\tilde{h}(f) \propto f^{-7/6} \Theta(f - 
f_\mathrm{ISCO})$, where the step function $\Theta$ is included to 
ensure that the radiation cuts off at $f_{\rm ISCO}$, the GW frequency at 
the innermost stable circular orbit of the binary. The 
post-Newtonian results are a weak field expansion and are only valid 
where velocities are much less than the speed of light. As a 
consequence, the leading-order post-Newtonian waveforms over-predict 
the S/N of an IMRI source, since they effectively spend too many 
cycles at each frequency as the ISCO is approached.

An alternative GW model can be obtained from perturbation theory, by 
expanding in terms of the mass ratio, $m/M$, assumed to be small. 
The IMRI systems considered in this paper lie somewhere between 
these two extremes --- the mass ratio is not quite small enough to 
use perturbative techniques, but the source spends a long time in 
the regime where post-Newtonian results are not valid. Waveform 
models have not yet been developed specifically for IMRI systems. 
However, by the time Advanced LIGO comes online, it is likely that 
models will have been constructed by combining post-Newtonian and 
perturbative techniques. This is discussed in more detail 
by~\cite{astrogr}. If accurate waveforms are not available, we will 
require sources to have higher S/Ns to be detected, thus reducing 
the ranges from the values that we quote. However, the loss in S/N 
from using an inaccurate template is likely to be only a few tens of 
percent~\citep{astrogr}, which is considerably smaller than the 
uncertainties in the astrophysical mechanisms that govern the event 
rates we are computing.

Out of the set of currently available waveform families, we believe 
that the most accurate S/Ns will come from the perturbative 
waveforms. Although the perturbative waveform will not be a precise 
model of the true waveform, the total energy content of the GWs will 
be roughly correct since the perturbative methods use a reliable 
model of the spacetime close to the central BH. To generate the 
range estimates quoted in this paper, we therefore computed the S/N 
via a perturbative model, as described below.

\citet{finnthorne00} used perturbation theory to compute the S/N, 
averaged over sky location and source orientation, contributed by the 
lowest four harmonics of the orbital frequency for circular, equatorial 
inspirals into Kerr BHs. Their calculation is accurate in the sense that 
it is based on perturbation theory, but it relies on three assumptions: 
(1) the orbit is in the extreme mass ratio limit, i.e., $m/M \ll 1$; (2) 
the orbit of the small body is circular; and (3) the orbit of the small 
body is equatorial. Assumption (2) is valid for our case, and assumption 
(1) is probably sufficiently accurate (the mass ratio here is 
intermediate while not extreme). Assumption (3) is not necessarily 
valid, but we can derive results for both prograde and retrograde 
equatorial orbits from the \citet{finnthorne00} waveforms and then 
average over possible orbital inclinations of the inspiraling object by 
assuming that the effect of averaging is the same as it is for the 
leading-order post-Newtonian model~\citep{Mandel}.

It is conventional to use $m$ to denote harmonic number when 
discussing harmonics of the azimuthal frequency. However, in this 
paper we use $k$ to avoid confusion with the mass of the CO.
The S/N contributed by the $k$th harmonic of the orbital frequency,
$f_k =k\omega_{orb}/(2\pi)$, 
and averaged over sky location and source orientation 
is given by~\citep{finnthorne00}
\begin{equation}
\rho_k^2 = \int{ \frac{\left[
h_{c,k}(f_k)\right]^2}{f_k S_n(f_k)}}d\ln f_k  ,
\label{SoverN}
\end{equation}
where $S_n(f)$ is the one-sided power spectral density of the detector
noise
and $h_{c,k}(f_k)$ is the characteristic amplitude of the $k$th harmonic
when
it passes through frequency $f_k$. This reduces to the earlier
expression~\erf{snrsq} via the substitution
$2\tilde{h}(f) = \sum_k h_{c,k} (f)/f$. The
characteristic amplitude is related to the energy radiated to infinity in
each
harmonic and is given by
\begin{eqnarray}
h_{c,1} = && \frac{5}{\sqrt{672\pi}} \frac{\sqrt{mM}}{r_o}
         \tilde\Omega^{1/6}{\cal H}_{c,1}, \label{hc1}\\ h_{c,k} =
         && \sqrt{\frac{5(k+1)(k+2)(2k+1)!k^{2k}} {12\pi (k-1)
         [2^k k! (2k+1)!!]^2}}  \times \frac{\sqrt{mM}}{r_o}
         \tilde\Omega^{(2m-5)/6}{\cal H}_{c,m} \quad \hbox{for~} m\ge2 .
\label{hcm} \end{eqnarray}
Here $\tilde{\Omega}=GM\omega_{orb}/c^3$ is the dimensionless orbital
angular frequency and $r_0$ is the distance to the source. The
relativistic
correction, ${\cal H}_{c,k}$, can be written as
\begin{equation}
{\cal H}_{c,k} = \sqrt{{\cal N}\dot {\cal E}_{\infty k}}.
\label{Hcm} \end{equation}
In this expression, ${\cal N}$ is the relativistic correction to the
number  of
cycles spent near a particular frequency, and ${\dot {\cal E}}_{\infty k}$
is
the relativistic correction to the rate of energy lost to infinity in
harmonic
$m$.  These corrections can be computed via integration of the
Teukolsky-Sasaki-Nakamura equations and are tabulated in
\cite{finnthorne00}.
We note that the various corrections are defined relative to their
Newtonian
values. 

Using the results of \citet{finnthorne00}, we can compute the total 
sky- and orientation-averaged S/N $\rho_{\rm tot}$ contributed by 
the lowest four harmonics of the orbital \ frequency from the time 
the source enters the detector band (when $f_4 = 10$Hz) until 
plunge, for various spins and masses of the central BH:
\begin{equation}
\rho_{\rm tot} = \sqrt{\sum_{k=1}^4 \rho^2_k}.
\end{equation}
This S/N was used to derive the range formulae presented in 
\S~\ref{LIGOrange}. We can also compute the leading-order 
post-Newtonian S/N by including only the quadrupolar $k=2$ mode and 
setting the correction ${\cal H}_{c,2} = 1$. We find that for $\chi 
\lesssim 0.5$ and $50\Msun < M < 250 \Msun$, the post-Newtonian S/N 
is typically an overestimate by a factor of $\sim 1.4$. We note that 
the data in \citet{finnthorne00} do not extend to the full range 
of radii needed for these calculations. Where necessary, we 
extrapolated their results to larger radius using appropriate power 
laws. We have verified that the results are insensitive to the exact 
form of this extrapolation.

The simplest template to use to detect a circular inspiral would 
include only the dominant, quadrupolar, component of the orbital 
frequency. It is useful to estimate how much S/N we would lose by 
ignoring higher harmonics. For circular inspirals in the equatorial 
plane of a Kerr BH, the fraction of the total energy radiated during 
an inspiral from infinity that is radiated between a certain 
Boyer-Lindquist radius $r_i$ and plunge effectively depends only on 
the ratio of the initial radius $r_i$ to the radius of the innermost 
stable circular orbit, $r_i/r_{\rm isco}(\chi)$, and is otherwise 
independent of $\chi$. Here $\chi$ is the central BH spin as usual, 
and $r_{isco}$ is the radius of the innermost stable circular orbit, 
as given in Eq.~\erf{iscorad}. The energy radiated in higher 
harmonics of the orbital frequency is suppressed relative to that in 
the dominant $k=2$ harmonic by powers of $M/r$. As the BH spin 
increases, $r_{\rm isco}/M\rightarrow1$ for prograde orbits, and so a 
larger fraction of the energy is radiated in the regime where $r\sim 
M$. We would therefore expect higher harmonics to contribute most 
significantly to the total energy flux for prograde inspirals into 
BHs with large spins. We computed the fraction of the total energy 
radiated into each harmonic as a function of the BH spin, while the 
particle inspirals from $r=10\,r_{\rm isco}$ to $r=r_{\rm isco}$. 
This is the range of radii for which \citet{finnthorne00} tabulate 
data, and in this range $\sim85\%$ of the total energy is radiated in 
any circular equatorial inspiral. The energy fractions are shown in 
Figure~\ref{Fig-energy}. We see that for $|\chi| \lesssim 0.3$, 
which is the expected IMBH spin range if the IMBH grows via minor 
mergers, $\sim 8\%$ of the energy is radiated into harmonics other 
than the dominant $k=2$ harmonic, and most of this energy goes into 
the $k=3$ harmonic.


The contribution of a harmonic to the S/N of a source depends not 
only on the energy that goes into that harmonic, but also on the 
shape of the noise curve --- higher harmonics enter the detector 
band earlier, contribute their signal at frequencies where the noise 
power spectral density is lower, and therefore have an enhanced 
contribution to the S/N. Figure~\ref{Fig-harmonics} shows the 
relative S/N contributed by each harmonic, defined as 
$\rho_k/\rho_{\rm tot}$, as a function of IMBH mass, for various 
IMBH spins. Note that this result does not depend on the mass of the 
inspiraling CO, since we are working in the extreme mass ratio 
limit. We see that for prograde inspirals, we can lose $\sim 
10-25\%$ of the S/N by using templates containing the $k=2$ mode 
only, but this is mostly recovered by including the $k=3$ mode in 
the search templates.  (We can lose up to $\sim 50\%$ of the S/N by 
using only $k=2$ templates for retrograde inspirals into high-mass 
IMBHs, but the S/Ns for such events are very small, making their 
detection unlikely.)


The S/Ns computed from these perturbative waveforms are not totally 
accurate for the reasons given earlier. Corrections will include 
finite-mass effects, contributions from the spin of the small BH and 
the effect of $k > 4$ harmonics of the orbital frequency. It is 
clear from Figure~\ref{Fig-energy} that for larger spins, a 
significant amount of energy goes into harmonics with $k>4$. These 
harmonics spend even longer in band and so their inclusion would 
increase the S/N. However, we cannot compute their contribution to 
the S/N since \cite{finnthorne00} do not tabulate these 
contributions separately. Overall, the S/Ns computed here should be 
accurate to $\sim 10\%$ and will be more accurate than those 
computed from the leading-order post-Newtonian waveforms.

\section{Ringdowns}
\label{ringdownapp}

Following the coalescence of an IMBH with a CO, the BH enters the 
ringdown phase, characterized by oscillations of its quasi-normal 
modes, particularly the dominant $l=m=2$ mode.  For IMRIs, the total 
energy emitted in GWs during the ringdown is $\sim 0.5 
m^2/M$~\citep{FH98}, which is a factor of $O(m/M)$ smaller than the 
total energy emitted over the inspiral.  However, the ringdown GW 
frequency~\citep{Echeverria},
\bel{fringdown}
f \approx \frac{1}{2\pi M} \left[ 1-0.63 (1-\chi)^{0.3} \right],
\ee
is higher than the ISCO frequency and is therefore closer to the 
minimum of the Advanced LIGO noise power spectral density for the 
typical masses under consideration.  For this reason, ringdowns may 
be detectable by Advanced LIGO despite their lower energy content.  
This is particularly true if $m \gtrsim 10\, M_\odot$ BHs, rather 
than NSs, are common as inspiraling companions, since the range for 
ringdowns scales as $m^2$ at low redshifts. Moreover, ringdowns will 
be the only way to detect CO coalescences with slowly-spinning IMBHs 
with masses above $350\, M_\odot$, since inspirals into such massive 
IMBHs will produce GWs at frequencies below the detector 
low-frequency limit.

\begin{figure}
\plotone{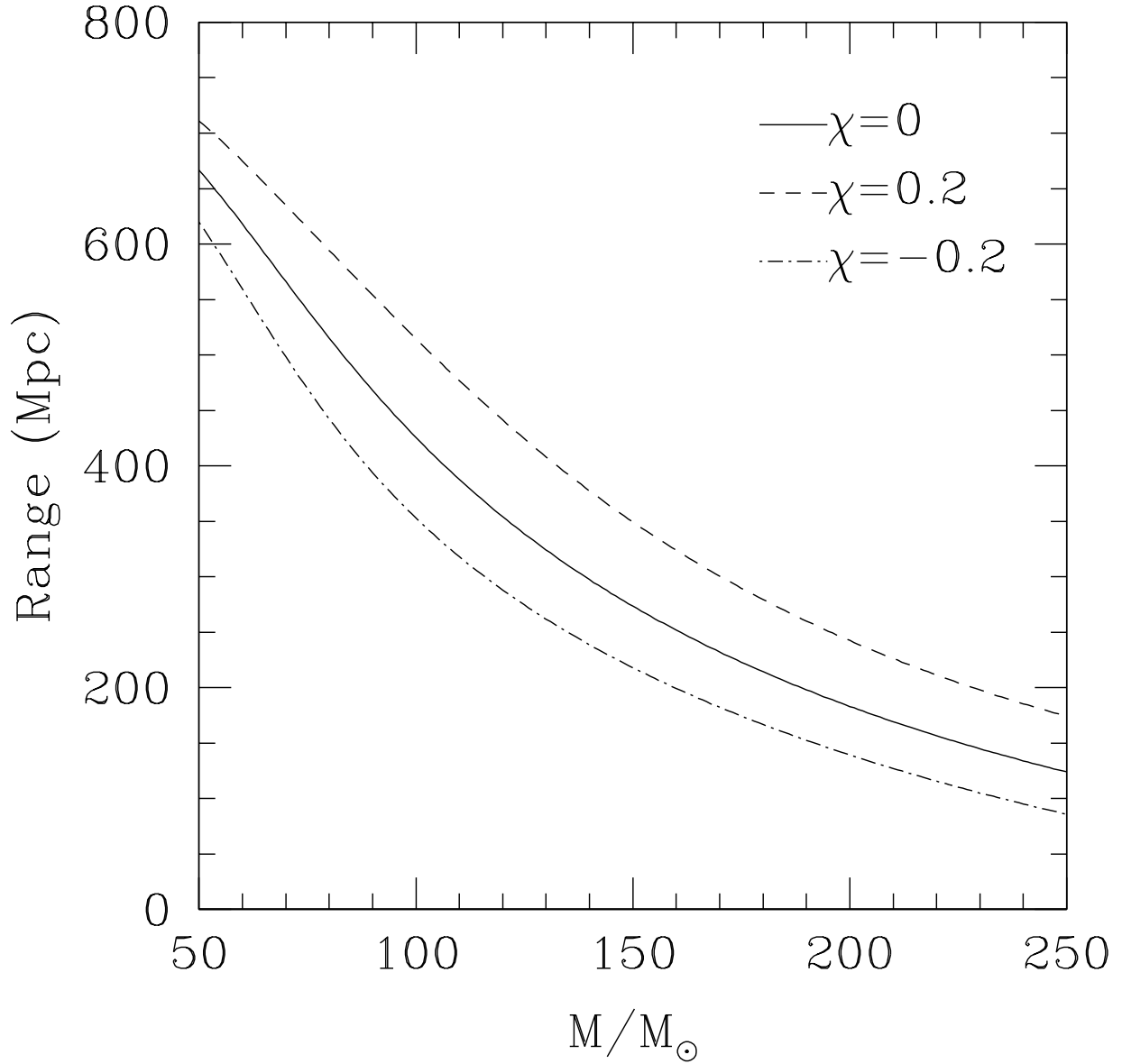}
\figcaption{Range of a network of three Advanced LIGO detectors for the
circular-equatorial-orbit inspiral of a $1.4\,M_\odot$ object into an
IMBH,
as a function of IMBH mass $M$. The three lines show IMRI spins of
$\chi = 0.2$ ({\it dashed}), $0$ ({\it solid}), and $-0.2$ 
({\it dot-dashed}).
Positive $\chi$ means prograde orbit; negative $\chi$ means retrograde.
The quadratic fit given in Eq.~(\ref{eq:rangefit}) is a fit to the
$\chi = 0$ line. \label{Fig-range}}
\end{figure}

\begin{figure}
\plotone{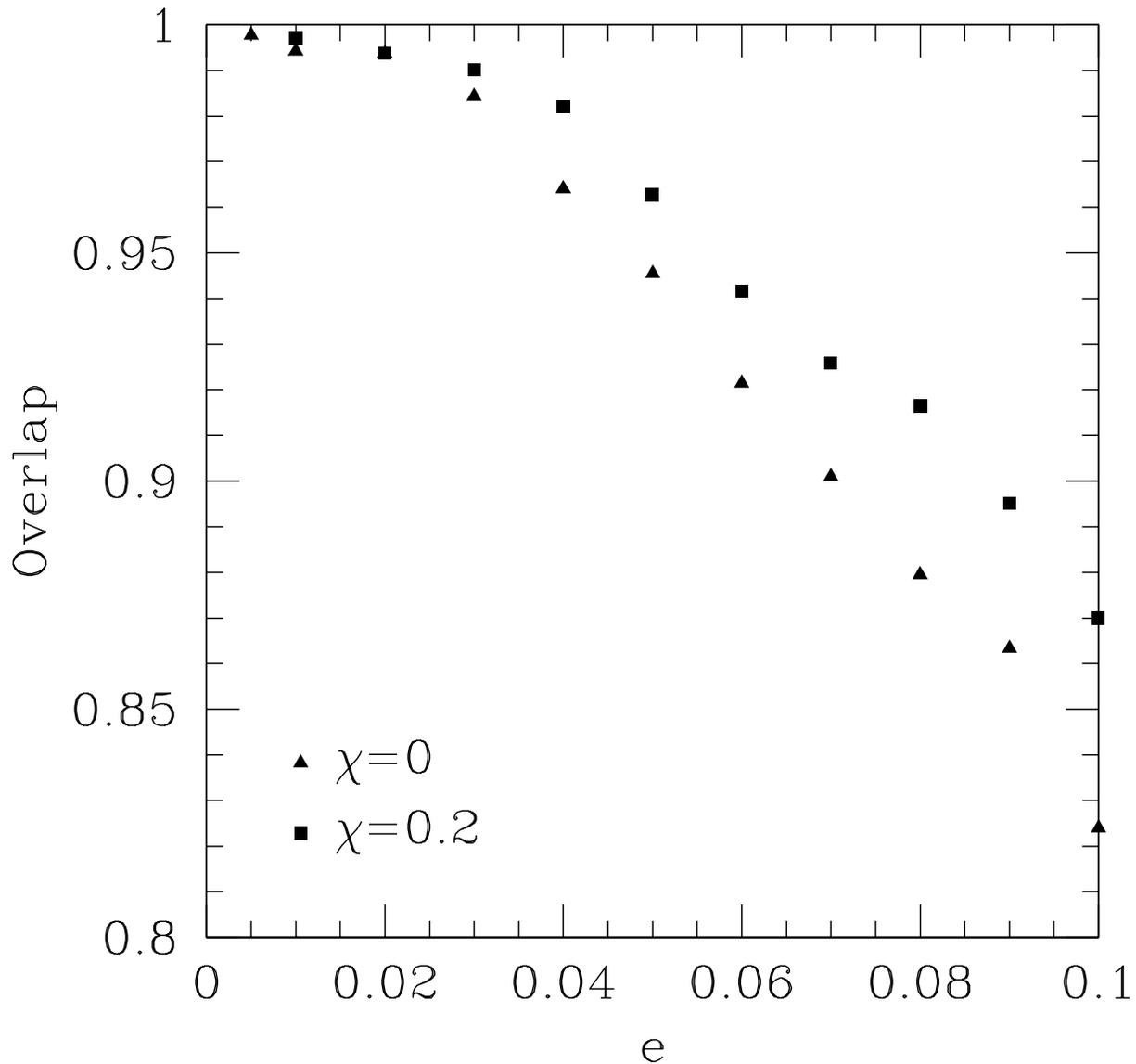}
\figcaption{Overlaps $\mathcal{O}$ between a circular template $h(t)$ and signals
$s(t)$ with varying eccentricities, $e$. For both signal and template,
the intrinsic parameters
$\vec{\theta} = (M = 100\, M_\odot, m = 1.4\,M_\odot, \chi, e)$ are
kept constant, with maximization performed only over time of arrival and
phase. The overlaps for two values of $\chi$ are shown.\label{f:overlap} }
\end{figure}

\begin{figure}
\plotone{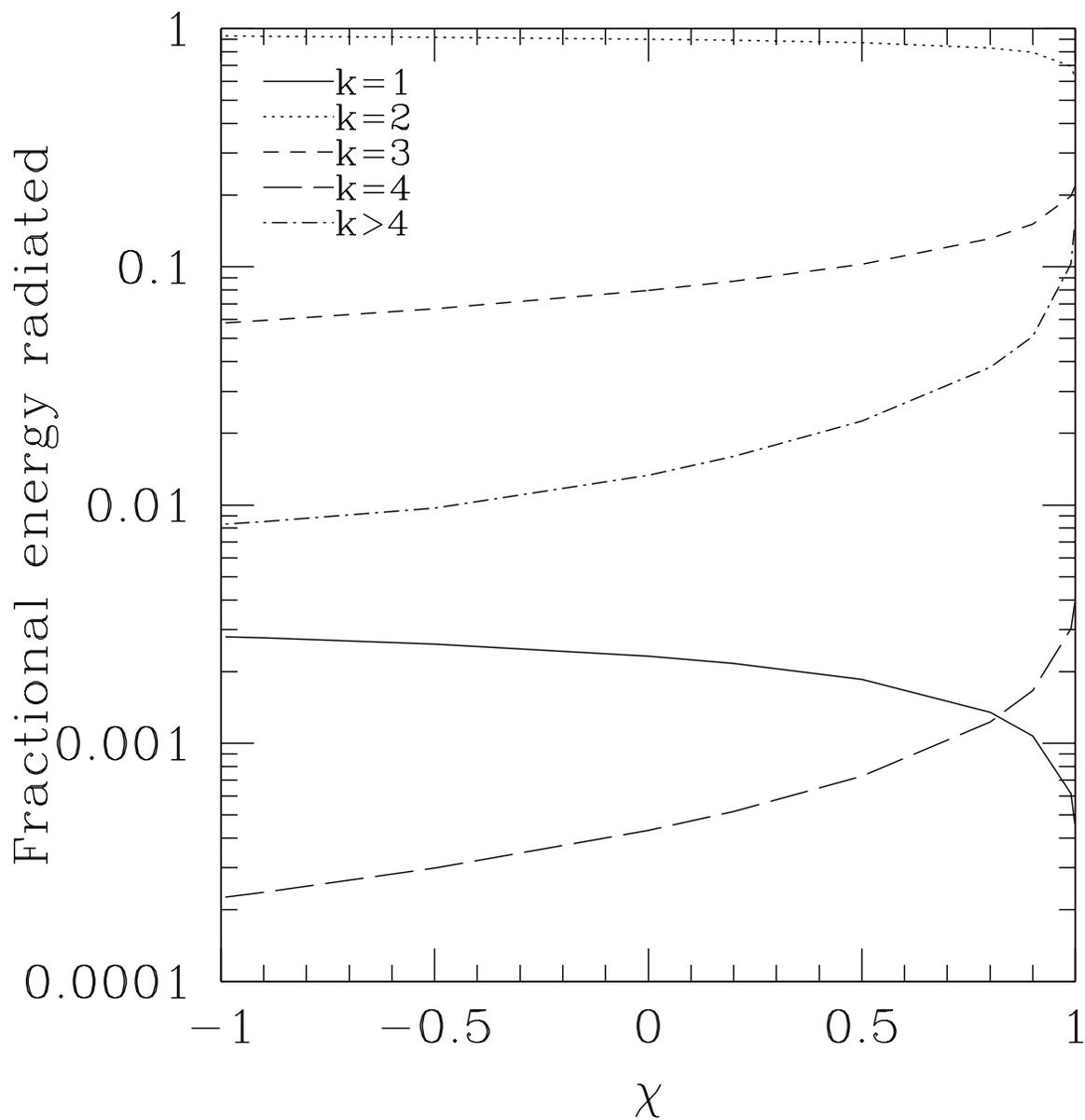}
\figcaption{Fraction of the total energy radiated into each harmonic of
the orbital frequency as the particle inspirals in a circular equatorial
orbit from  $10\,r_{\rm isco}$ to $r_{\rm isco}$. This energy fraction is
shown as a function of BH spin.\label{Fig-energy}}
\end{figure}

\begin{figure}
\plotone{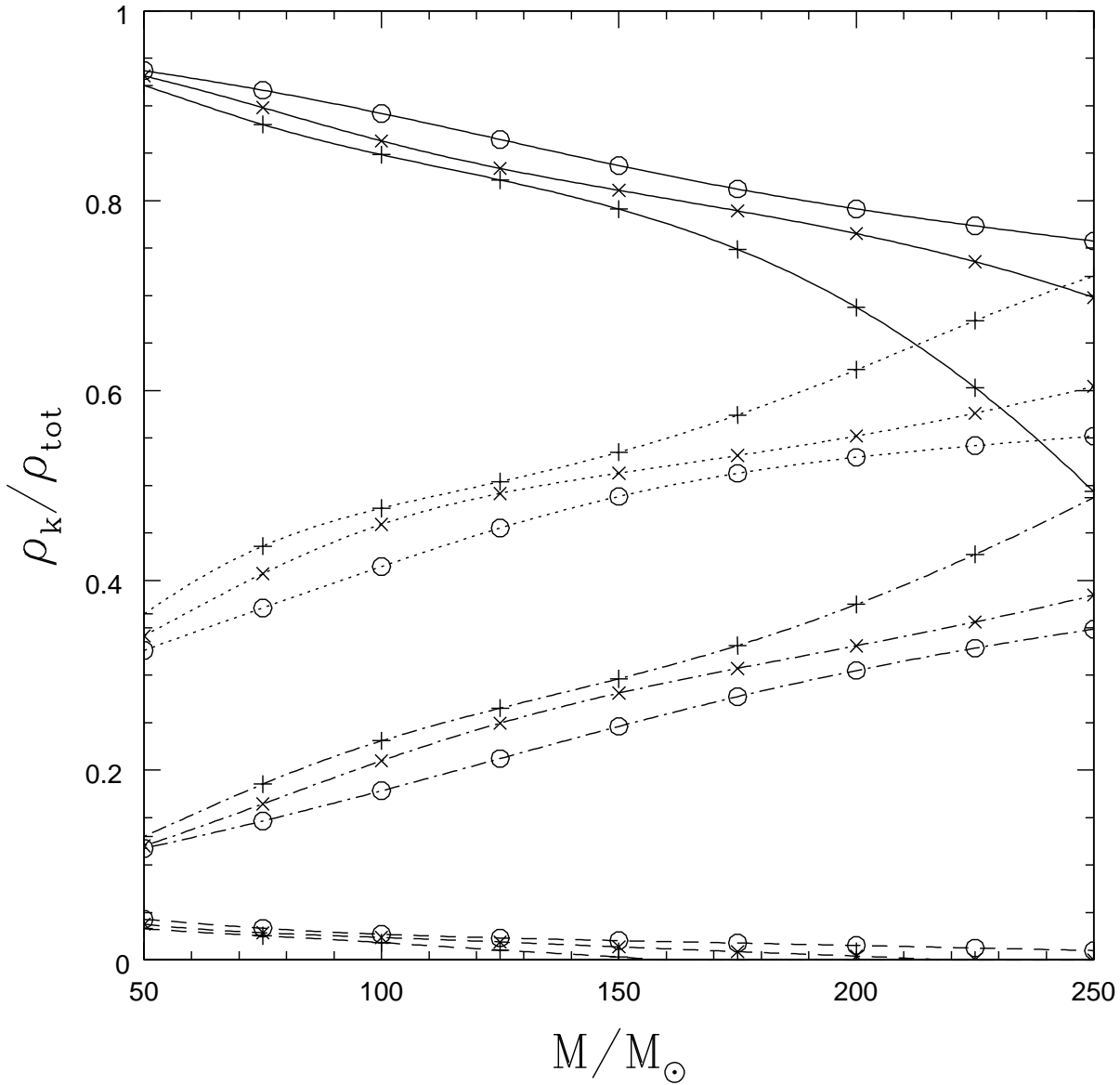}
\figcaption{Ratio of sky- and orientation-averaged S/N contributed by 
the lowest four harmonics of the orbital frequency to the total sky- and 
orientation-averaged S/N contributed by the lowest four harmonics, 
as a function of the central BH mass, for circular equatorial orbits.
The harmonics are indicated by
different line styles --- $k=1$ ({\it dashed}), $k=2$ ({\it solid}), 
$k=3$ ({\it dotted}),
and
$k=4$ ({\it dot-dashed}). Lines are shown for three different BH spins,
$\chi = 0$, $\chi = 0.5$ and $\chi = -0.5$ (i.e., retrograde inspirals
into
a $\chi=0.5$ BH), indicated by crosses,
circles, and plus signs, respectively.\label{Fig-harmonics}}
\end{figure}

\begin{figure}
\plotone{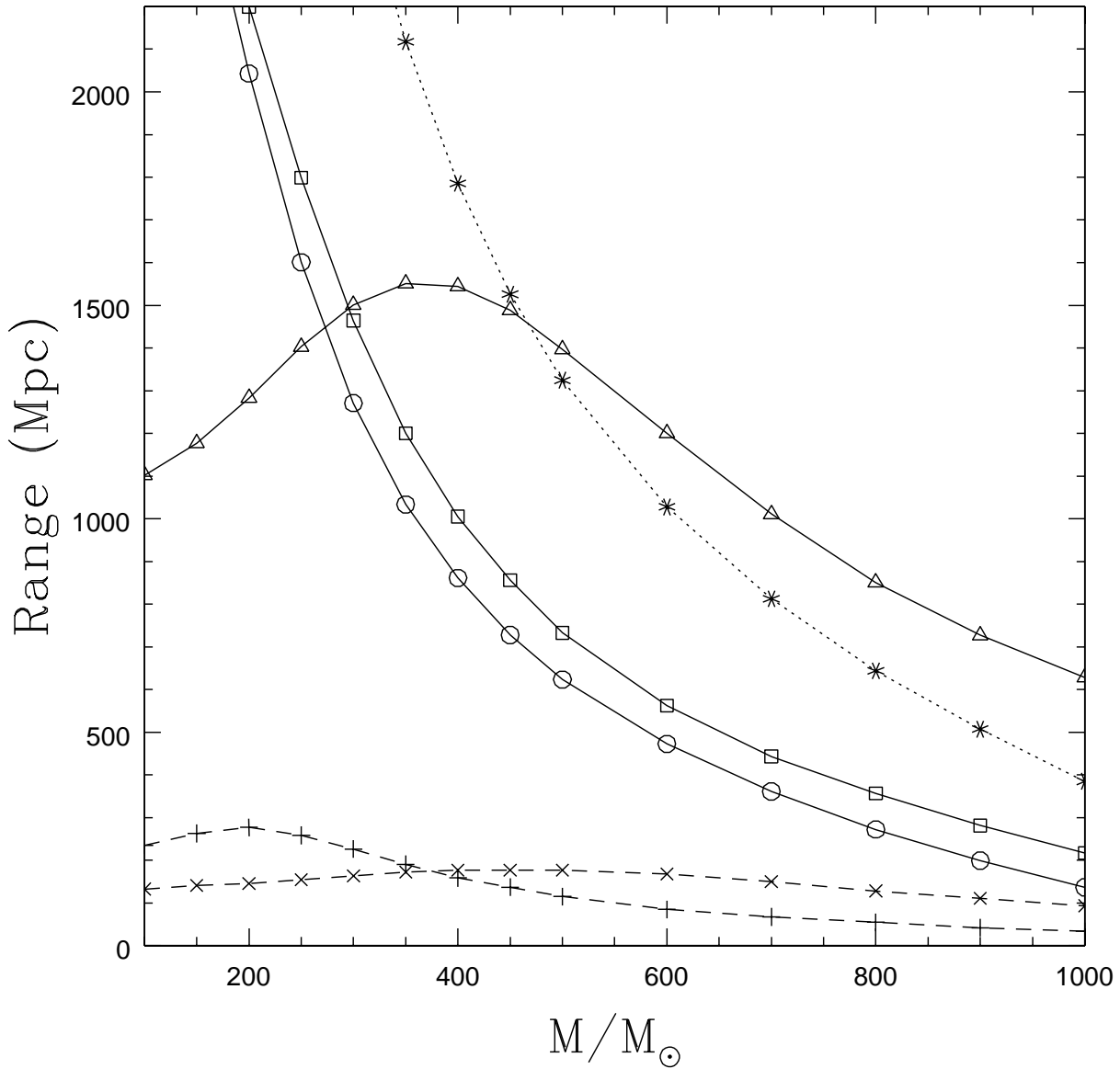}
\figcaption{Range of a network of three Advanced LIGO detectors  for the
ringdown of an IMBH following a merger with a CO.
The luminosity-distance range in Mpc is plotted as a function of
IMBH mass $M$;  cosmological redshift is included.  Dashed lines denote
$m=1.4\, M_\odot$ inspiraling NSs, with plus signs 
corresponding to IMBH spin
$\chi=0.3$ and crosses to $\chi=1$.  Solid lines denote $m=10\, M_\odot$
inspiraling  BHs, with circles, squares, and triangles corresponding to
spins $\chi=0$, $\chi=0.3$, and $\chi=1$, respectively.  The 
dotted line with asterisks
denotes $m=20\, M_\odot$ BHs spiraling into an IMBH with spin
$\chi=0.3$.\label{Fig-ringdown}}
\end{figure}

The typical Advanced LIGO ringdown-wave ranges (in terms of 
luminosity distance) as a function of IMBH mass are plotted in 
Fig.~\ref{Fig-ringdown} for several choices of inspiraling object 
mass and IMBH spin.  Because some ranges reach out to significant 
redshifts (up to $z \sim 0.5$), the effect of redshifting is already 
included in these ranges, unlike in Fig.~\ref{Fig-range}.  
Redshifting also explains why the range does not scale strictly as 
$m^2$, as high redshifts bring the GW frequency at the detector down 
into the region where the interferometer is less sensitive.


The astrophysical rate of ringdowns per cluster is greater than or 
equal to the rate of IMRIs, since every IMRI culminates in a merger 
and ringdown (but ringdowns could follow coalescences without 
observable inspirals, i.e., those with direct plunges).  The 
distance sensitivity to ringdowns following inspirals of $1.4\, 
M_\odot$ NSs is probably too low to make them detectable by Advanced 
LIGO: the total detectable event rate for NS--IMBH ringdowns is 
$\sim 20$ times lower than the event rate for NS--IMBH inspirals if 
the IMBH mass is $M=100\, M_\odot$ and spin is $\chi=0.3$.  However, 
Advanced LIGO will be considerably more sensitive to ringdowns than 
to inspirals in other mass ranges.  For example, ringdowns from $10\ 
M_\odot + 300\ M_\odot$ coalescences could be detected in a volume 
$\sim 200$ times greater than the detection volume for inspirals 
from these coalescences; if all IMBHs had mass $M = 300 M_\odot$, 
and all COs were $m=10\ M_\odot$ BHs with coalescence rate equal to 
$\approx 5 \times 10^{-9}$ yr$^{-1}$ per cluster as in 
\S~\ref{harden}, then the total detectable IMRI ringdown event rate 
would reach $\sim 20$ yr$^{-1}$ . Thus, if our expectations about the 
likely masses involved in IMRIs are incorrect, and coalescences of 
COs with higher masses with more massive IMBHs are common, searches 
for ringdown waves can provide a useful back-up to IMRI searches.

\clearpage

\begin{thebibliography}{}

\bibitem[Abbott et al.(2008)]{S3S4}
{Abbott}, B., et al. (LIGO Scientific Collaboration). 2008, \prd, 77, 062002 
 
\bibitem[Allen et al.(2005)]{Allen:2005fk}
{Allen}, B., {Anderson}, W.~G., {Brady}, P.~R., {Brown}, D.~A., \&
         {Creighton}, J.~D.~E. 2005, arXiv:gr-qc/0509116

\bibitem[Amaro-Seoane et al.(2007)]{astrogr}
{Amaro-Seoane}, P., {Gair}, J.~R., {Freitag}, M., {Miller}, M.~C., {Mandel}, I., {Cutler}, C.~J. \& {Babak}, S. 2007, \cqg, 24, R113

\bibitem[Apostolatos(1996)]{Apostolatos}
{Apostolatos}, T.~A. 1996, \prd, 52, 605

\bibitem[Babak et al.(2007)]{babak06}
{Babak}, S.~V., {Fang}, H., {Gair}, J.~R., {Glampedakis},  K., \&
{Hughes},
S.~A. 2007, \prd, 75, 024005

\bibitem[Bahcall \& Wolf(1976)]{BahcallWolf}
{Bahcall}, J.~N., \& {Wolf}, R.~A. 1976, \apj, 209, 214

\bibitem[{{Baker} {et al.}(2007)}]{Baker07}
{Baker}, J.~G., {Boggs}, W.~D., {Centrella}, J., {Kelly}, B.~J.,
{McWilliams}, S.~T., {Miller}, M.~C., \& {van Meter}, J. 2007,
ApJ, submitted (arXiv:astro-ph/0702390)

\bibitem[Baker et al.(2006)]{Baker06}
{Baker}, J.~G., Centrella, J., Choi, D.-I., Koppitz, M., van Meter, J.~R.,
\& Miller, M.~C. 2006, \apj, 653, L93

\bibitem[Bardeen, Press, \& Teukolsky(1972)]{BPT72}
{Bardeen}, J.~M., {Press}, W.~H., \& {Teukolsky}, S.~A. 1972, \apj, 178,
347

\bibitem[Barish \& Weiss(1999)]{Barish:1999}
{Barish}, B.~C., \& {Weiss}, R. 1999, Phys.~Today, 52, 44

\bibitem[Bekenstein(1973)]{Bek73}
{Bekenstein}, J.~D. 1973, \apj, 183, 657

\bibitem[Bellazzini et al.(2002)]{bellazzini02}
{Bellazzini}, M., {Fusi Pecii}, F., {Montegriffo}, P., {Messineo}, M., {Monaco}, L. \& {Rood}, R.~T. 2002, Astron.~J., 123, 1509

\bibitem[Blanchet, Qusailah, \& Will(2005)]{BQW05}
{Blanchet}, L., {Qusailah}, M.~S.~S., \& {Will}, C.~M. 2005, \apj, 635,
508

\bibitem[Brown et al.(2007)]{Brown}
{Brown}, D.~A., Brink, J., Fang, H., Gair, J.~R., Li, C., Lovelace, G.,
Mandel,
I., \& Thorne, K.~S. 2007, \prl, 99, 201102

\bibitem[{{Campanelli} {et al.}(2007a)}]{Campanelli07a}
{Campanelli}, M., {Lousto}, C.~O., {Zlochower}, Y., \&
{Merritt}, D. 2007a, \apj, 659, L5

\bibitem[{{Campanelli} {et al.}(2007b)}]{Campanelli07b}
{Campanelli}, M., {Lousto}, C.~O., {Zlochower}, Y., \&
{Merritt}, D. 2007b, \prl, 98, 231102

\bibitem[Cutler \& Flanagan(1994)]{Cutler:1994ys}
{Cutler}, C., \& {Flanagan}, E.~E. 1994, \prd, 49, 2658

\bibitem[Damour \& Gopakumar(2006)]{DG06}
{Damour}, T., \& {Gopakumar}, A. 2006, \prd, 73, 124006

\bibitem[Ebisuzaki et al.(2001)]{Ebisuzaki01}
{Ebisuzaki}, T., et al. 2001, \apj, 562, L19

\bibitem[Echeverria(1988)]{Echeverria}
{Echeverria}, F. 1988, \prd, 40, 3194

\bibitem[Favata, Hughes, \& Holz(2004)]{FHH04}
{Favata}, M., {Hughes}, S.~A., \& {Holz}, D.~E. 2004, \apj, 607, L5

\bibitem[Finn(1987)]{finn87}
{Finn}, L.~S. 1987, MNRAS, 227, 265

\bibitem[Finn \& Chernoff(1993)]{Finn:1992xs}
{Finn}, L.~S., \& {Chernoff}, D.~F. 1993, \prd, 47, 2198

\bibitem[Finn \& Thorne(2000)]{finnthorne00}
{Finn}, L.~S., \& Thorne, K.~S. 2000, \prd, 62, 124021

\bibitem[Fitchett(1983)]{Fit83}
{Fitchett}, M.~J. 1983, MNRAS, 203, 1049

\bibitem[Fitchett \& Detweiler(1984)]{FD84}
{Fitchett}, M.~J., \& {Detweiler}, S. 1984, MNRAS, 211, 933

\bibitem[Flanagan \& Hughes(1998)]{FH98}
{Flanagan}, E.~E., \& {Hughes}, S.~A. 1998, \prd, 57, 4535

\bibitem[Flanagan \& Racine(2007)]{FR07}
{Flanagan}, E.~E., \& {Racine}, E. 2007, \prd, 75, 044001

\bibitem[Fregeau et al.(2006)]{Fregeau06}
{Fregeau}, J.~M., {Larson}, S.~L., {Miller}, M.~C.,
{O'Shaughnessy}, R., \& {Rasio}, F.~A. 2006, \apj, 646, L135

\bibitem[Freitag(2003)]{freitag02}
{Freitag}, M., 2003, ApJ, 583, L21

\bibitem[Freitag, G\"urkan, \& Rasio(2006)]{FGR06}
{Freitag}, M., {G\"urkan}, M.~A., \& {Rasio}, F.~A. 2006,
MNRAS, 368, 141

\bibitem[Freitag, Rasio, \& Baumgardt(2006)]{FRB06}
{Freitag}, M., {Rasio}, F.~A., \& {Baumgardt}, H. 2006, MNRAS,
368, 121

\bibitem[Friedman \& Schutz(1978)]{friedman78}
{Friedman}, J.~L., \& Schutz, B.~F. 1978, \apj, 221, 937

\bibitem[Fritschel(2003)]{Fritschel}
Fritschel, P. 2003, arXiv:gr-qc/0308090

\bibitem[Fryer \& Kalogera(2001)]{FK2001}
{Fryer}, C.~L., \& {Kalogera}, V. 2001, \apj, 554, 548

\bibitem[Gair \& Glampedakis(2006)]{GG06}
{Gair}, J.~R., \& {Glampedakis}, K. 2006, \prd, 73, 064037

\bibitem[Gonzalez et al.(2007a)]{Gonzalez06}
{Gonzalez}, J.~A., Sperhake, U., Bruegmann, B., Hannam, M., \& Husa, S.
2007a, \prl, 98, 091101

\bibitem[{{Gonzalez} {et al.}(2007b)}]{Gonzalez07}
{Gonzalez}, J.~A., {Hannam}, M.~D., {Sperhake}, U., {Br\"ugmann}, B.,
\& {Husa}, S. 2007b, \prl, 98, 231101 


\bibitem[G\"ultekin, Miller, \& Hamilton(2004)]{GMH04}
{G\"ultekin}, K., {Miller}, M.~C., \& {Hamilton}, D.~P. 2004,
ApJ, 616, 221

\bibitem[G\"ultekin, Miller, \& Hamilton(2006)]{GMH06}
{G\"ultekin}, K., {Miller}, M.~C., \& {Hamilton}, D.~P. 2006,
ApJ, 640, 156

\bibitem[G\"urkan, Fregeau, \& Rasio(2006)]{GFR06}
{G\"urkan}, M.~A., {Fregeau}, J.~M., \& {Rasio}, F.~A. 2006,
ApJ, 640, L39

\bibitem[G\"urkan, Freitag, \& Rasio(2004)]{GFR04}
{G\"urkan}, M.~A., {Freitag}, M., \& {Rasio}, F.~A. 2004, ApJ,
604, 632

\bibitem[Heggie (1975)]{Heggie75}
{Heggie}, D.~C. 1975, MNRAS, 173, 729

\bibitem[Heggie, Trenti, \& Hut(2006)]{Heggie06}
{Heggie}, D.~C., {Trenti}, M., \& {Hut}, P. 2006, MNRAS, 368, 677

\bibitem[{{Herrmann} {et al.}(2007a)}]{Herrmann06}
{Herrmann}, F., {Hinder}, I., Shoemaker, D., \& Laguna, P. 2007a, \cqg, 24, 33

\bibitem[{{Herrmann} {et al.}(2007b)}]{Herrmann07}
{Herrmann}, F., {Hinder}, I., {Shoemaker}, D., {Laguna}, P.,
\& {Matzner}, R.~A. 2007b, \apj, 661, 430

\bibitem[Ho \& Lai(1999)]{Ho99}
{Ho}, W.~C.~G. \& {Lai}, D. 1999, MNRAS, 308, 153

\bibitem[Hopman \& Alexander(2005)]{HA05}
{Hopman}, C., \& Alexander, T. 2005, ApJ, 629, 362

\bibitem[Hopman \& Portegies Zwart(2005)]{HPZ05}
{Hopman}, C., \& Portegies Zwart, S.~F. 2005, MNRAS Lett., 363, L56

\bibitem[Hopman, Portegies Zwart, \& Alexander(2004)]{HPZA03}
{Hopman}, C., Portegies Zwart, S.~F., \& Alexander, T. 2004, ApJ, 604,
L101

\bibitem[Hughes \& Blandford(2003)]{HB03}
{Hughes}, S.~A., \& {Blandford}, R.~D. 2003, ApJ, 585, L101

\bibitem[Hurley(2007)]{Hurley07}
{Hurley}, J.~R. 2007, MNRAS, 379, 93

\bibitem[Innanen et al.(1997)]{I97}
{Innanen}, K.~A., {Zheng}, J.~Q., {Mikkola}, S., \& {Valtonen}, M.~J.
1997.
AJ, 113 (5), 1915

\bibitem[{{Koppitz} {et al.}(2007)}]{Koppitz07}
{Koppitz}, M., {Pollney}, D., {Reisswig}, C., {Rezzolla}, L.,
{Thornburg}, J., {Diener}, P., \& {Schnetter}, E. 2007,
arXiv:gr-qc/0701163

\bibitem[Kozai(1962)]{K62}
{Kozai}, Y. 1962, AJ, 67, 591

\bibitem[Kulkarni, Hut, \& McMillan(1993)]{KHM93}
{Kulkarni}, S.~R., Hut, P., \& McMillan, S.~L.~W.
1993, Nature, 364, 421

\bibitem[Madau \& Rees(2001)]{MR01}
{Madau}, P., \& Rees, M.~J. 2001, ApJ, 551, L27

\bibitem[Mandel(2007)]{Mandel}
{Mandel}, I. 2007, ApJ, submitted, arXiv:0707.0711

\bibitem[Martel \& Poisson(1999)]{Martel99}
{Martel}, K., \& Poisson, E. 1999, \prd, 60, 124008

\bibitem[Miller (2002)]{M02}
{Miller}, M.~C. 2002, ApJ, 581, 438

\bibitem[Miller \& Colbert(2004)]{MC04}
{Miller}, M.~C., \& {Colbert}, E.~J.~M. 2004, IJMPD, 13, 1

\bibitem[Miller \& Hamilton(2002a)]{MH02a}
{Miller}, M.~C., \& {Hamilton}, D.~P. 2002a, MNRAS, 330, 232

\bibitem[Miller \& Hamilton(2002b)]{MH02b}
{Miller}, M.~C., \& {Hamilton}, D.~P. 2002b, ApJ, 576, 894

\bibitem[Mouri \& Taniguchi(2002a)]{MT02a}
{Mouri}, H., \& {Taniguchi}, Y. 2002a, ApJ, 566, L17

\bibitem[Mouri \& Taniguchi(2002b)]{MT02b}
{Mouri}, H., \& {Taniguchi}, Y. 2002b, ApJ, 580, 844

\bibitem[O'Leary, O'Shaughnessy, \& Rasio(2007)]{OL07}
{O'Leary}, R., {O'Shaughnessy}, R., \& {Rasio}, F.~A. 2007, PRL, submitted
         (arXiv:astro-ph/0701887)

\bibitem[O'Leary et al.(2006)]{OL06}
{O'Leary}, R.~M., {Rasio}, F.~A., {Fregeau}, J.~M., {Ivanova}, N.,
\& {O'Shaughnessy}, R. 2006, ApJ, 637, 937

\bibitem[Peres(1962)]{Per62}
{Peres}, A. 1962, Phys.~Rev., 128, 2471

\bibitem[Peters(1964)]{P64}
{Peters}, P.~C. 1964, \prb, 136, 1224

\bibitem[Peters \& Mathews(1963)]{Peters:1963ux}
{Peters}, P.~C., \& {Mathews}, J. 1963, Phys.~Rev., 131, 435

\bibitem[Phinney(1991)]{Phinney91}
{Phinney}, E.~S. 1991, ApJ, 380, L17

\bibitem[Phinney(2005)]{Phinney}
{Phinney}, E.~S. 2005, private communication

\bibitem[Portegies Zwart et al.(2004)]{PZ04}
{Portegies Zwart}, S., {Baumgardt}, H., {Hut}, P.,
{Makino}, J., \& {McMillan}, S.~L.~W. 2004, Nature, 428, 724

\bibitem[Portegies Zwart \& McMillan(2000)]{PZ}
{Portegies Zwart}, S., \& McMillan, S.~L.~W. 2000, ApJ, 528, L17

\bibitem[Portegies Zwart \& McMillan(2002)]{PM02}
{Portegies Zwart}, S., \& {McMillan}, S.~L.~W. 2002, ApJ, 576, 899

\bibitem[Press \& Teukolsky(1977)]{press77}
{Press}, W.~H., \& Teukolsky, S.~A. 1977, ApJ, 213, 183

\bibitem[Pryor \& Meylan(1993)]{pryor93}
{Pryor}, C., \& Meylan, G. 1993, in {\it Structure and Dynamics
of Globular Clusters.}, eds.~Djorgovski S.~G., Meylan G.
(San Francisco: ASP), vol.~50, p.~357

\bibitem[Quinlan(1996)]{Q96}
{Quinlan}, G.~D. 1996, New Astronomy 1, 35

\bibitem[Quinlan \& Shapiro(1987)]{QS87}
{Quinlan}, G.~D., \& Shapiro, S.~L. 1987, ApJ, 321, 199

\bibitem[Quinlan \& Shapiro(1989)]{QS89}
{Quinlan}, G.~D., \& {Shapiro}, S.~L. 1989, ApJ, 343, 725

\bibitem[Redmount \& Rees(1989)]{RR89}
{Redmount}, I.~H., \& {Rees}, M.~J. 1989, Commun.~Astrophys., 14, 165

\bibitem[Reisenegger \& Goldreich(1992)]{reisen92}
{Reisenegger}, A., \& Goldreich, P. 1992, ApJ, 395, 240

\bibitem[Rubenstein \& Bailyn(1997)]{rubenstein97}
{Rubenstein}, E.~P., \& Bailyn, C.~D. 1997, ApJ, 474, 701

\bibitem[Sigurdsson \& Hernquist(1993)]{SH93}
{Sigurdsson}, S., \& Hernquist L. 1993, Nature, 364, 423

\bibitem[Sopuerta, Yunes, \& Laguna(2007)]{Sop06}
{Sopuerta}, C.~F., Yunes, N., \& Laguna, P. 2007, \apj, 656, L9

\bibitem[Taniguchi et al.(2000)]{Tan00}
{Taniguchi}, Y., {Shioya}, Y., {Tsuru}, T.~G., \& {Ikeuchi},
S. 2000, PASJ, 52, 533

\bibitem[Thorne(1987)]{thorne.k:1987}
{Thorne}, K.~S., in
  \emph{Three hundred years of gravitation}, edited by
  S.~W.~Hawking and W.~Israel
  (Cambridge University Press, Cambridge, 1987),
  chap.~9, pp.~330--458

\bibitem[Trenti(2006)]{Trenti06b}
{Trenti}, M. 2006, arXiv:astro-ph/0612040

\bibitem[Trenti et al.(2007)]{Trenti06a}
{Trenti}, M., {Ardi}, E., {Mineshige}, S., \& {Hut},P. 2007, MNRAS, 374, 857

\bibitem[Trenti, Heggie, \& Hut(2007)]{Trenti07}
{Trenti}, M., {Heggie}, D.~C., \& {Hut}, P. 2007, MNRAS, 374, 344

\bibitem[Vallisneri(2000)]{vallis00}
Vallisneri, M. 2000, \prl, 84, 3519

\bibitem[Wainstein \& Zubakov(1962)]{Wainstein:1962}
{Wainstein}, L.~A., \& {Zubakov}, V.~D. 1962,
         "Extraction of signals from noise", Prentice-Hall,
         Englewood Cliffs, NJ

\bibitem[Wen(2002)]{Wen02}
{Wen}, L. 2002, ApJ, 598, 419

\bibitem[Wiseman(1992)]{W92}
{Wiseman}, A.~G. 1992, \prd, 46, 1517

\end{thebibliography}
\end{document}